\documentclass[journal]{IEEEtran}

\usepackage{setspace}
\usepackage{lettrine}

\usepackage{graphics,
	psfrag,
	epsfig,
	amsthm,
	cite,
	amssymb,
	url,
	dsfont,
	algorithmic,
	balance,
	enumerate,
	color,
	setspace
}
\usepackage{amsmath}

\usepackage{epstopdf}

\newcommand{\eref}[1]{(\ref{#1})}
\newcommand{\sref}[1]{Section~\ref{#1}}

\newcommand{\cref}[1]{Constraint~\ref{#1}}

\newcommand{\ignore}[1]{}

\usepackage[labelformat=simple]{subcaption}

\IEEEoverridecommandlockouts
\usepackage{mathtools}
\usepackage{color}
\usepackage[utf8]{inputenc}
\usepackage[ruled, lined, longend]{algorithm2e}
\SetEndCharOfAlgoLine{}

\begin{document}
\IEEEoverridecommandlockouts


\title{\vspace{-.9cm} Energy Efficient Federated Learning in Integrated
	Fog-Cloud Computing Enabled  Internet-of-Things Networks}

\author{
 \IEEEauthorblockN{Mohammed S. Al-Abiad, \textit{Student Member, IEEE}, Md. Zoheb Hassan, \textit{Student Member, IEEE}, and Md. Jahangir Hossain, \textit{Senior Member, IEEE}}

\thanks {
Mohammed S. Al-Abiad and Md. Jahangir Hossain are with the School of
	Engineering, University of British Columbia, Kelowna, BC V1V 1V7, Canada
	(e-mail: m.saif@alumni.ubc.ca, jahangir.hossain@ubc.ca).
	
	Md. Zoheb Hassan is with $\acute{\text{E}}$cole de technologie sup$\acute{\text{e}}$rieure (ETS), University of Quebec, Canada (e-mail:  md-zoheb.hassan.1@ens.etsmtl.ca).
}
\vspace{-0.4cm}
}

\maketitle
\begin{abstract}
We investigate resource allocation scheme  to reduce the energy consumption of federated learning (FL) in the integrated fog-cloud computing enabled Internet-of-things (IoT) networks. In the envisioned system, IoT devices are connected with the centralized cloud server (CS) via multiple fog
access points (F-APs). We consider two different scenarios for training the local models. In the first
scenario, local models are trained at the IoT devices and the F-APs upload the local model
parameters to the CS. In the second scenario, local models are trained at the F-APs based on the collected data from the IoT devices and the F-APs collaborate with the CS for updating the model parameters. Our objective is to minimize the overall energy-consumption of both scenarios subject to FL time constraint. Towards this goal, we devise a joint optimization of scheduling of IoT devices with the F-APs, transmit power allocation, computation frequency allocation at the devices and F-APs and decouple it into two subproblems. In the
first subproblem, we optimize the IoT device scheduling and power allocation, while in the second subproblem, we optimize the computation frequency allocation. For each scenario, we develop a conflict graph based solution to
iteratively solve the two subproblems. Simulation results show that the
proposed two schemes achieve a considerable performance
gain in terms of the energy consumption minimization. The presented simulation results interestingly reveal that for a large number of IoT devices and large data sizes, it is more energy efficient to train the local models at the IoT devices instead of the F-APs. 

\end{abstract}


\section{Introduction} \label{sec:I}
The emerging Internet-of-Things (IoT) leads to the unprecedented growth of the connected IoT devices in the wireless networks and significant rise  of several computation demanding applications, such as interactive gaming, virtual/augmented reality, image/video processing.  Cloud computing provides an efficient computation platform for executing the aforementioned applications. However, cloud computing requires efficient  offloading of computation intensive tasks from the energy-constrained mobile devices to the cloud server (CS) of enormous computation capability \cite{CC2}. In particular, the offloading of a task to a distant CS increases latency and security risk (e.g., important data should not be offloaded to CSs that are located outside a national territory) \cite{CC3}. Hence, the benefit of cloud computing is diminished for the latency-sensitive and  security critical applications \cite{CC4}. Fog computing provides a complementary solution to the contemporary cloud computing by reducing the distance between the computing CS and mobile devices \cite{FC1}. 

In fog computing systems, fog access points (F-APs), with certain storage and computation capabilities, are deployed at the network edge \cite{FC2, 8,FC_Zoheb_1,FC_Zoheb_2}. As a result,  mobile devices can offload the computation intensive tasks directly to the nearby F-APs instead of a distant CS, leading to the low-latency and fast access services. Moreover, an integration of fog and cloud computing provides a powerful computation and communication platform for the large number of IoT devices. Such an integrated architecture is referred as integrated fog-cloud computing (FCC)-enabled IoT system \cite{FCC_1}. The integrated FCC provides powerful computation architecture for the IoT device, and it benefits from  the centralized signal processing at the CS. 
Thereby, the  integrated FCC can efficiently provide the required quality-of-service (QoS) for the emerging IoT applications.

Recently,  the data driven decision making becomes an integral part of IoT networks, thanks to the availability of the enormous data and advancement of the devices' computing power. In particular, machine learning algorithms are extensively used to predict traffic congestion, user behavior, and QoS of users by analyzing large-scale data collected from the IoT devices.   In conventional ML,  the collected data from IoT devices (e.g., images, videos, and recorded audios) are offloaded to and processed in the centralized CSs, where
the learning models are trained.  Such a centralized ML approach is confronted by the huge traffic burden in the wireless links between IoT devices and core network. In addition, the privacy of users' sensitive data is impeded.
Therefore, the conventional centralized learning method is inefficient for next generation  IoT network \cite{Distributed_learning}.  Federated learning (FL) has been emerged as an efficient decentralized learning mechanism that allows multiple network edge devices to collaboratively learn a shared model \cite{FL_0}.  In FL, the network edge devices train  learning models locally based on local data. In contrast to the centralized learning mechanism,  the devices only share updated model parameters with the CS. Subsequently, the CS calculates the global model parameter by aggregating  local model updates from edge network. The local and global parameters are updated iteratively until convergence. By distributing the learning tasks between the network edge and centralized CSs, FL not only reduces the  huge traffic burden over wireless channel, but also protects  privacy of IoT data \cite{FL_1}.

Due to the availability of distributed computing resources, an integrated FCC system provides a convenient platform to implement FL in wireless networks. In fact, using FL in integrated FCC system, the QoS of users can be significantly improved \cite{FL_2}. However,  the channel impairments and interference present the key challenges to implement FL over wireless networks. Specifically, the FL training loss and convergence time jointly depends on selection of the collaborating  devices, spectrum resource allocation, power allocation, and computation capability of the collaborating devices \cite{FL_3}. It is also imperative to reduce the energy consumption of the IoT devices participating in FL process. Particularly, IoT devices consume both communication and computation energy when the local models are trained at the IoT devices. To save the computation energy consumption,  fog computing resources can be leveraged for local learning.  However, in  such a scenario, the communication energy consumption can be increased as the IoT devices need to upload a large amount of data to nearby F-APs with a strict latency constraint.  To this end, we consider two different scenarios where the local models are trained either at the IoT devices or F-APs. For both scenarios, we develop resource allocation mechanisms to reduce energy consumption of FL with strict delay constraints.

\subsection{Related Works and Motivations}

\textit{Related works on communication-efficient FL:}   The  performance of a decentralized ML  depends on the optimization of wireless links between the network edge devices and  parameter server (CS or fog computing nodes). Hence, it is imperative to optimally design the learning-centric  resource allocation schemes \cite{FL1}. In the recent literature, the design of communication-efficient FL was extensively studied. Leveraging the grouping of  network edge devices and a decentralized group alternating direction method of multipliers, a jointly communication efficient and fast converging FL algorithm was proposed in \cite{FL_Com_1}. To enhance the accuracy and convergence of FL, it is imperative to enhance the number of collaborating edge devices while using the available spectrum resources efficiently. To this end, a collaborative FL framework was proposed that allows resource constrained IoT devices to upload model parameters to the nearby devices instead of the distance CS \cite{FL_Com_2}. Moreover, a joint scheduling of network edge devices and radio resource blocks (RRBs) was studied to minimize the FL loss function via applying Lyapunov optimization framework \cite{FL_Com_3}.  In a heterogeneous cellular network, a hierarchical FL framework  can effectively  enhance the number of devices participating in local learning \cite{FL_Com_4}. In such a hierarchical FL framework,  at each round, the network edge devices upload their model parameters only to the nearest F-APs (therein called small base-stations), and F-APs periodically upload the average local model parameters  to the CS (therein called macro base-station) for a global aggregation.  Thus, a large number of devices can participate in local learning.  Besides, interference among the network edge devices  can induce error in FL and increase the convergence time. Accordingly, interference aware radio resource allocation is also imperative for communication-efficient FL framework. A joint optimization of user selection, RRB allocation, and transmit power allocation was presented to minimize the loss function in FL training process. The authors in \cite{FL_Com_5} proposed transmit power allocation of the  IoT devices to enhance information freshness in FL system. Considering the presence of eavesdroppers in an Internet-of-drones network, the authors in \cite{7a} proposed a secured and delay-constrained FL scheme through transmit power allocations. 

\textit{Related works on energy-efficient FL:}   Since the mobile devices are battery-driven, for a sustainable operation of an FL framework, it is imperative to reduce energy consumption of the edge devices. In particular, an energy-efficient or green FL should consider minimizing communication and computation energy. In \cite{EE_FL_1}, energy-efficient radio resource allocation was proposed for delay constrained FL. However, the authors in \cite{EE_FL_1} only minimized the communication energy and ignored the computation energy. In \cite{EE_FL_2}, the authors proposed an adaptive FL framework, where the devices can send quantized or compressed model parameters and thus, save energy. However, the radio resource optimization was not presented in \cite{EE_FL_2}. In \cite{EE_FL_3}, radio resource allocation was developed to minimize both communication and computation energy in an FL system subject to delay constraints. However, the authors in \cite{EE_FL_3} considered orthogonal multiple access (OMA)  to connect edge devices with the base-station, which can limit the number of collaborating devices.  Using OMA, the authors in \cite{FCR11} also proposed joint transmit power and computation frequency allocation to reduce overall energy consumption of FL in a fog-aided IoT network. The energy limitation of the collaborating edge devices can  also be improved by energy-harvesting technique \cite{EE_FL_4}. Moreover, the work in \cite{EE_FL_5} considered a game theory framework to motivate the network edge devices to participate in local learning while reducing its energy consumption.

\textit{Motivations and Challenges:}  In contrast to the existing works \cite{EE_FL_1,EE_FL_2,EE_FL_3,FCR11,EE_FL_4,EE_FL_5}, our motivation  is to develop resource allocation mechanisms to facilitate energy-efficient FL in an integrated FCC-enabled IoT network. The considered architecture has a number of F-APs along with an CS, and the IoT devices are connected with the CS through F-APs.  To improve the number of connected devices with F-APs using limited RRBs, an uplink non-OMA (NOMA) scheme \cite{NOMA5} is considered. To the best of our knowledge, this is the first work that investigates efficient integration of joint cloud-fog computing and NOMA technique to reduce energy consumption of FL scheme. However,  to take advantage of such an architecture for reducing energy consumption of FL, we need to develop a computationally efficient resource allocation scheme. Specifically, we need to address the following two challenges.

\begin{itemize}
	\item \textbf{Challenge I:} In the first scenario, F-APs can  work as relays where all the local learning is executed at the IoT devices. Alternatively, F-APs have the computation capability, and thus, they can participate in local learning along with the IoT devices. Therefore, in the second scenario, F-APs can work  as local learning nodes to save the computation energy consumption of the connected IoT devices.  Essentially, we need to investigate which of the aforementioned two scenarios has better energy-efficiency.
	
	\item \textbf{Challenge II:}   There is an inherent trade-off between FL time and energy consumption. In particular, both computation and communication energy are increased to reduce the FL time.  Accordingly, to satisfy a given FL time constraint and reduce energy consumption, it is imperative to jointly optimize the degrees-of-freedom, namely, power allocation, IoT device scheduling to the F-APs/RRBs, and computation frequency allocation. However, an interplay of the aforementioned factors leads to a high computational complexity. Essentially, we need to develop a computationally efficient resource allocation scheme to address the trade-off between FL time and energy consumption.

\end{itemize}

\subsection{Contributions} 
We investigate resource allocation  for energy-efficient FL in an integrated FCC-enabled IoT network. Specifically, we propose a joint optimization of scheduling of IoT devices with the F-APs/RRBs, transmit power allocations, and computation frequency allocation at the IoT devices and F-APs. To this end, we introduce innovative graph-theoretical frameworks to develop  computationally efficient solution.  The main contributions of our work are as follows. 
\begin{enumerate}
	\item We consider two different scenarios for training the local model. In the first scenario, referred as  \textit{IoT device local learning}, local models are trained at the IoT devices and the F-APs upload the collected local model parameters to the CS for aggregation. In the second scenario,  referred as \textit{F-AP local learning}, the local models are trained at the F-APs based on the collected data from the IoT devices.
	For both scenarios, we aim at minimizing the overall energy consumption of IoT devices and F-APs subject to FL time constraint, IoT device-RRB/F-AP scheduling, transmit power allocations, and computation frequency allocation. Such a joint optimization problem is NP-hard and thus, computationally intractable. By analyzing the problem of each scenario, we decompose it into two subproblems namely, \textit{resource scheduling and power
		allocation} subproblem and \textit{computation frequency allocation} subproblem. 

	\item To solve the first subproblem,  using graph theory, we design a low-complexity algorithm to optimize the scheduling among the IoT devices, F-APs, and RRBs.  and transmit power levels of the IoT devices. In contrast, we obtain closed-form computation frequency allocation solution that depends on the scheduling  obtained in the first subproblem. For both scenarios, efficient solutions are obtained by solving the aforementioned  two subproblems  alternately. The computational complexities of the proposed schemes are analyzed as well. 
	
	%
	%
	%
	
	\item Extensive
	simulations are conducted to verify advantages of the proposed schemes over the benchmark schemes. Numerical  results revealed that both proposed
	schemes offer improved energy consumption performances as compared to the benchmark schemes. The presented simulation results also interestingly reveal that for a large number of IoT devices and large data sizes, it is more energy efficient to train the local models at the IoT devices instead of the F-APs. 
\end{enumerate}

The rest of this paper is organized as follows. The system model is described  in  \sref{SMMM}.  The optimization problems for the considered scenarios are provided in \sref{PF}. In \sref{G} and \sref{MT}, we develop two graph theory schemes to facilitate local learning at the IoT devices and F-APs, respectively. Simulation results are presented in
\sref{NR}, and in \sref{C}, we conclude the paper.

\section{System Overview} \label{SMMM}
\ignore{\begin{figure}[t!]
	\centering
	\includegraphics[width=0.55\linewidth]{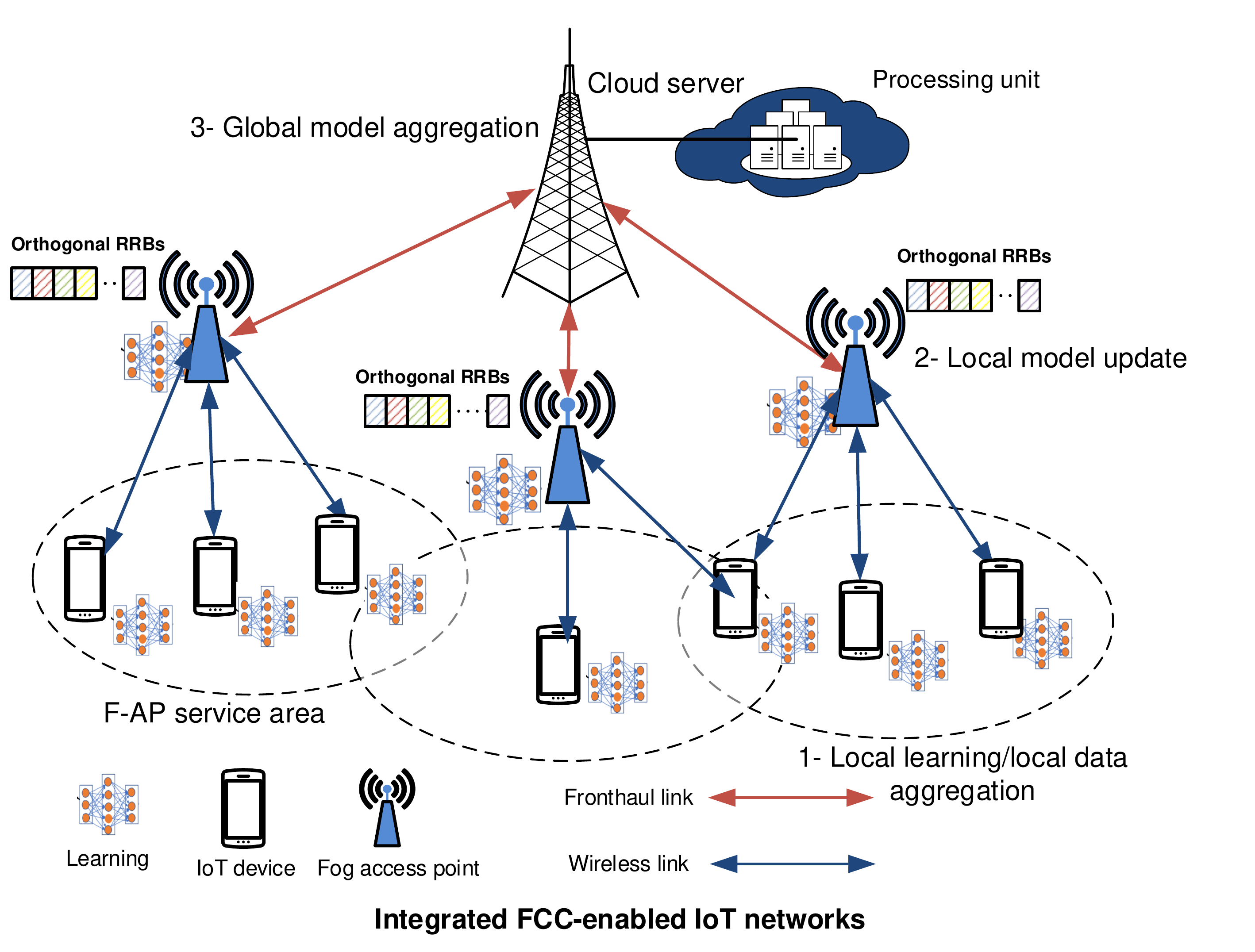}
	\caption{Illustration of integrated 	FCC-enabled IoT networks.}
	\label{fig1}
\end{figure}}

\subsection{System Model}
We consider an integrated
FCC-enabled IoT system, illustrated in Fig. \ref{fig1}, that consists of one cloud server (CS), $K$ F-APs, and $N$ IoT devices. The sets of IoT devices and F-APs are denoted
by  $\mathcal{N}=\{1,2,\cdots,N\}$ and  $\mathcal{K}=\{1,2,\cdots,K\}$, respectively. The $N$ IoT devices  (e.g., smartphones, laptops, and cameras) are connected to the F-APs which are connected  to the CS using fronthaul links. We consider that each F-AP has a limited coverage range that represents the service area of the $k$-th F-AP within a circle of radius $\mathtt R$. The set of IoT devices in the $k$-th F-AP's coverage range is defined by $\mathcal N_{k}=\{n\in \mathcal{N}| d_{k,n}\leq \mathtt R$\}, where $d_{k,n}$ is the distance between the $k$-th F-AP and the $n$-th IoT device. Let $\mathbf A={a_{k,n}}$ be
the F-AP allocation matrix, where element $a_{k,n}=1$ represents that the $n$-th IoT device is allocated to the $k$-th F-AP, and $a_{k,n}=0$ otherwise.
\ignore{The IoT devices independently train local ML models based on their aggregated local data (e.g., images, videos, recorded audios), and upload model parameters to the nearest F-APs, and then, the F-APs receive all the model parameters and upload them to the CS for aggregation. }

\ignore{\begin{figure}[t!]
	\centering
	\begin{subfigure}[t]{0.44\textwidth}
		\centerline{\includegraphics[width=.8501\linewidth]{fig_system_model_new.pdf}}
		\label{fig1}
		
	\end{subfigure}%
	~
	\begin{subfigure}[t]{0.44\textwidth}
		\centerline{\includegraphics[width=0.85703\linewidth]{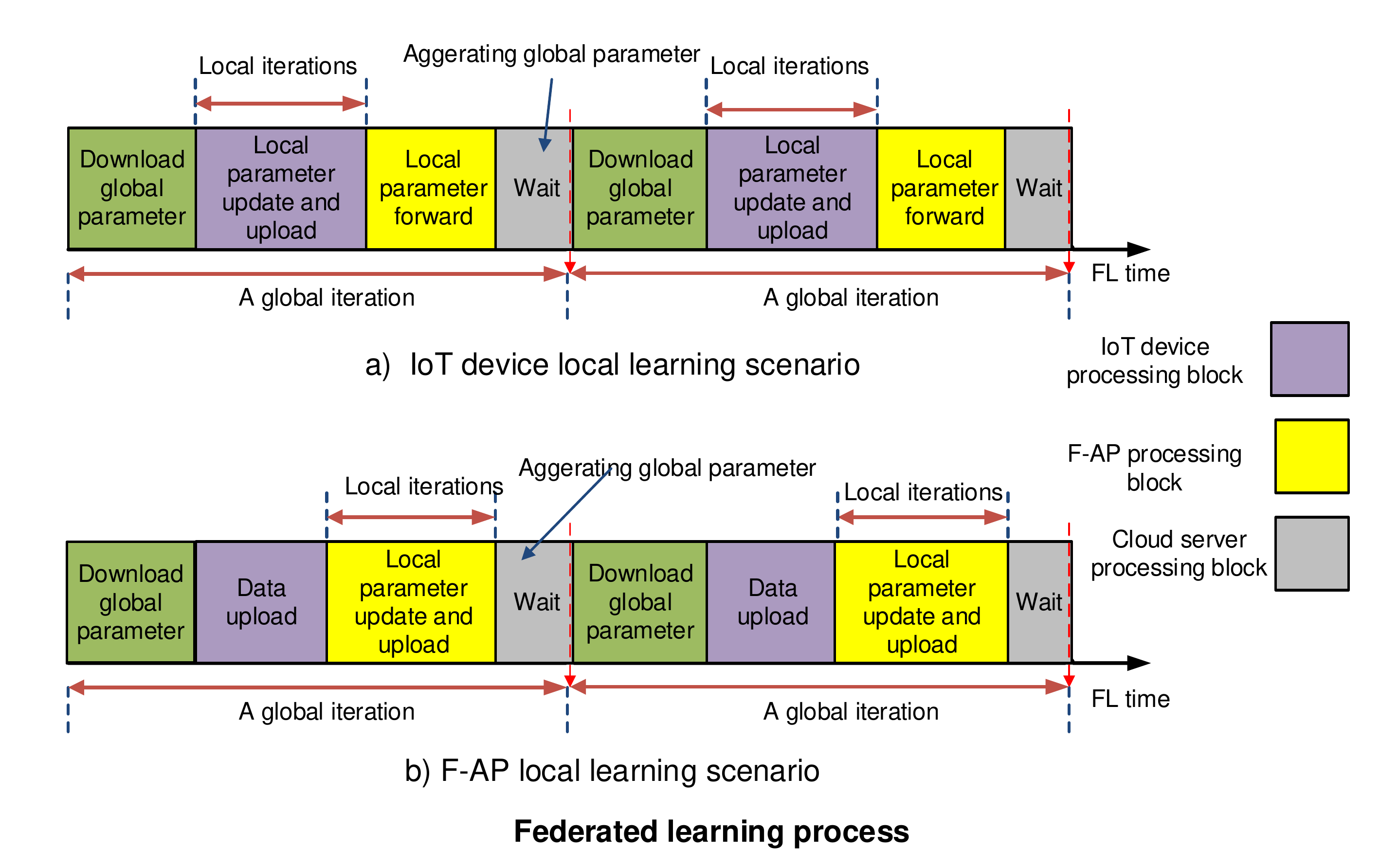}}
		\label{NCT}
	\end{subfigure}
	
	\caption{Illustration of integrated 	FCC-enabled IoT networks and FL process.}
	\label{fig1}
\end{figure}}

\begin{figure}[t!]
	\centering
	\includegraphics[width=0.85\linewidth]{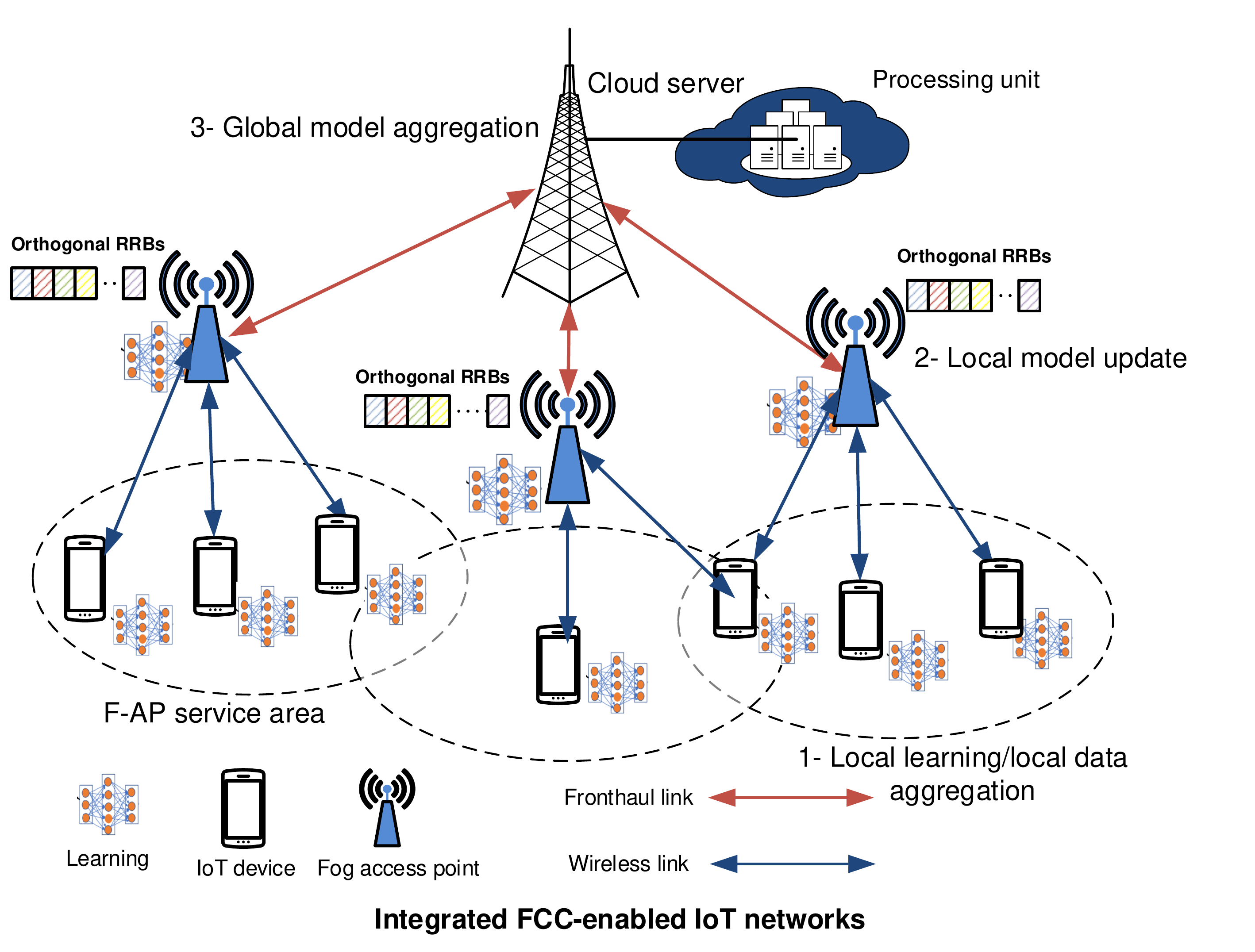}
	\caption{Illustration of integrated 	FCC-enabled IoT networks}
	\label{fig1}
\end{figure}

Let $\mathcal D_n$ denote the local data set of IoT device $n$, which is a set of data samples $\{x_i, y_i\}$, where $x_i$ is sample $i$’s input (e.g., image
pixels) and $y_i$ is sample $i$’s output (e.g., label of the image). Similar to \cite{EE_FL_2, FCR11},
the local loss function on IoT device $n$’s data set can be calculated as $L_n(\omega)=\frac{1}{D_n}\sum_{i\in \mathcal D_n}l_i(\omega), \forall n\in \mathcal N,$ where $D_n=|\mathcal D_n|$ is the number of collected data samples by IoT device $n$ and $l_i(\omega)$ is the loss function that measures the local training model error of data sample $i$. Then, IoT device $n$ finds the optimum $\omega^*_n$ that minimizes $L_n(\omega)$ and uploads it to the suitable F-APs for aggregation by the CS. The IoT devices independently train local ML models based on their aggregated local data (e.g., images, videos, recorded audios). As shown in
Fig. \ref{fig1}, the specific process of FL  in the $t$-th iteration can be summarized as: 1) each IoT
device downloads the global model parameters $\omega_n(t-1)$ from
the CS through the nearest F-AP; 2) each IoT device updates the local model by its
local training data and sends the updated local
model parameter $\omega_n(t)$ back to the F-APs; and 3) the CS aggregates the information from the F-APs and calculates the
new global model parameters. 

Each IoT device uploads the local information to the nearest F-AP via a wireless link. Similar to the resource setting in \cite{8a, 9}, we consider that each F-AP has $Z$ orthogonal RRBs that are denoted by the set $\mathcal{Z}=\{1,2,\cdots,Z\}$, where IoT devices can transmit their local information to the F-APs. These RRBs can be used practically as a generic term to denote time/frequency resource block of
every F-AP, i.e., a group of  orthogonal sub-carriers \cite{8}. Let $\mathbf S=\{s^n_{k,z}\}$ be the RRB allocation matrix, where element $s^n_{k,z}=1$ represents that the $n$-th IoT device is allocated to the $k$-th F-AP on the $z$-th RRB, and $s^n_{k,z}=0$ otherwise. In this work,
we consider a simple and efficient system’s design where the scheduling-level coordination takes
place, i.e., each user is only scheduled to a single RRB \cite{8, 8a, 9}. To schedule number of IoT devices to each RRB, we consider NOMA. Let $p_n$ denote the transmission power of the $n$-th IoT device and let $\bf p$ be a $1\times N$ matrix
containing the power levels of all IoT devices, i.e., 
$\textbf p = [p_n]$. Hence, the instantaneous signal-to-interference-plus-noise (SINR)  for the link between the $n$-th IoT device and the $z$-th RRB in the $k$-the F-AP is given by
\begin{align} 
\label{SNR}
\gamma^n_{k,z}=\frac{s^n_{k,z}p_n \left|G^n_{k,z}\right|^2}{\sum_{\substack{j \in \mathcal{N}_k\backslash n\\G^{j}_{k,z}<G^n_{k,z}}}s^{j}_{z,k}p_{j} \left|G^{j}_{k,z}\right|^2 +\sigma^2}, \forall (j,n) \in \mathcal N_k,
\end{align} 
where $\sigma^{2}$ denotes the additive white Gaussian noise variance and $G^n_{k,z}$ denotes the channel gain for the link between the $n$-th IoT device and the $z$-th RRB in the $k$-th F-AP. Then, the transmit rate of the $n$-th IoT device to the $k$-th F-AP over the $z$-th RRB can be given by $R^n_{k,z}=W\log_{2}(1+\gamma^n_{k,z})$, where $W$ is the bandwidth of the $z$-th RRB. Consequently, the transmit rate of the $n$-th IoT device  is $R_n=\sum_{k\in \mathcal K}\sum_{z\in \mathcal Z}W\log_{2}(1+\gamma^n_{k,z})$. At the F-APs, each F-AP uploads its collected local parameters to the CS through a fronthaul
link of capacity $R_{fh}$.

\begin{figure}[t!]
	\centering
	\includegraphics[width=0.85\linewidth]{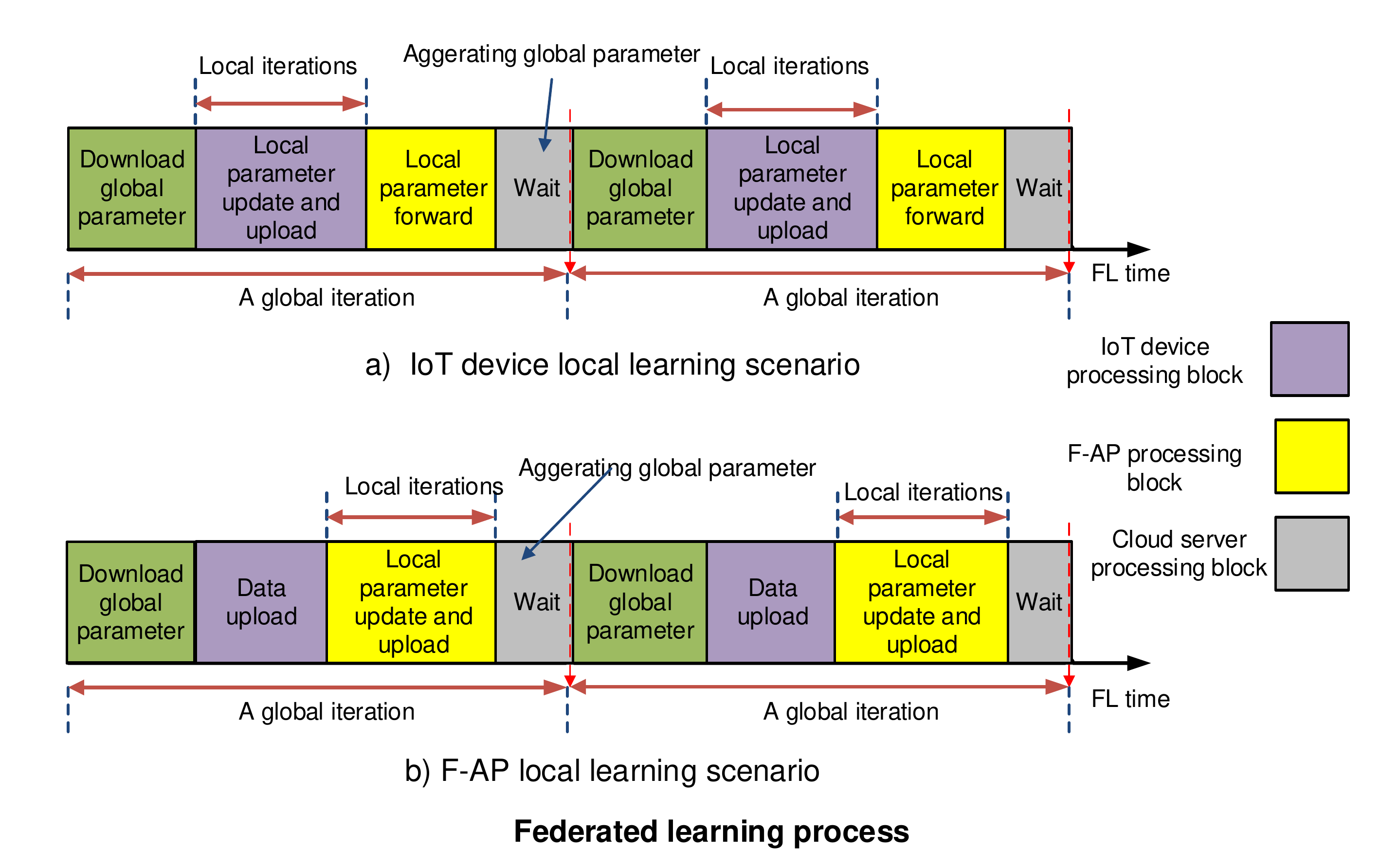}
	\caption{Illustration of federated learning process.}
	\label{fig2a}
\end{figure}

\ignore{\begin{figure}[t!]
	\centering
	\begin{subfigure}[t]{0.44\textwidth}
		\centerline{\includegraphics[width=.8501\linewidth]{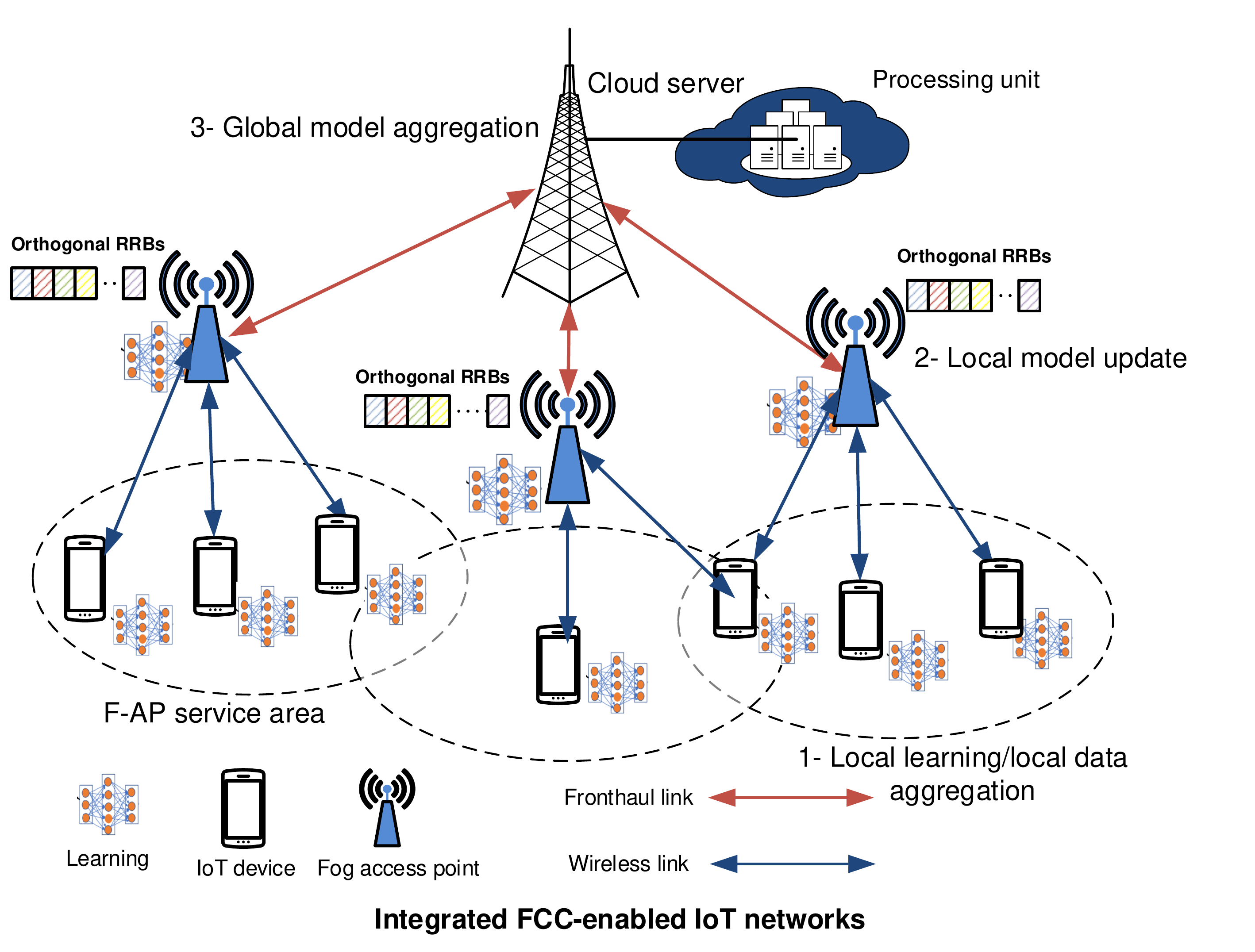}}
		\label{fig1}
		
	\end{subfigure}%
	~
	\begin{subfigure}[t]{0.44\textwidth}
		\centerline{\includegraphics[width=0.85703\linewidth]{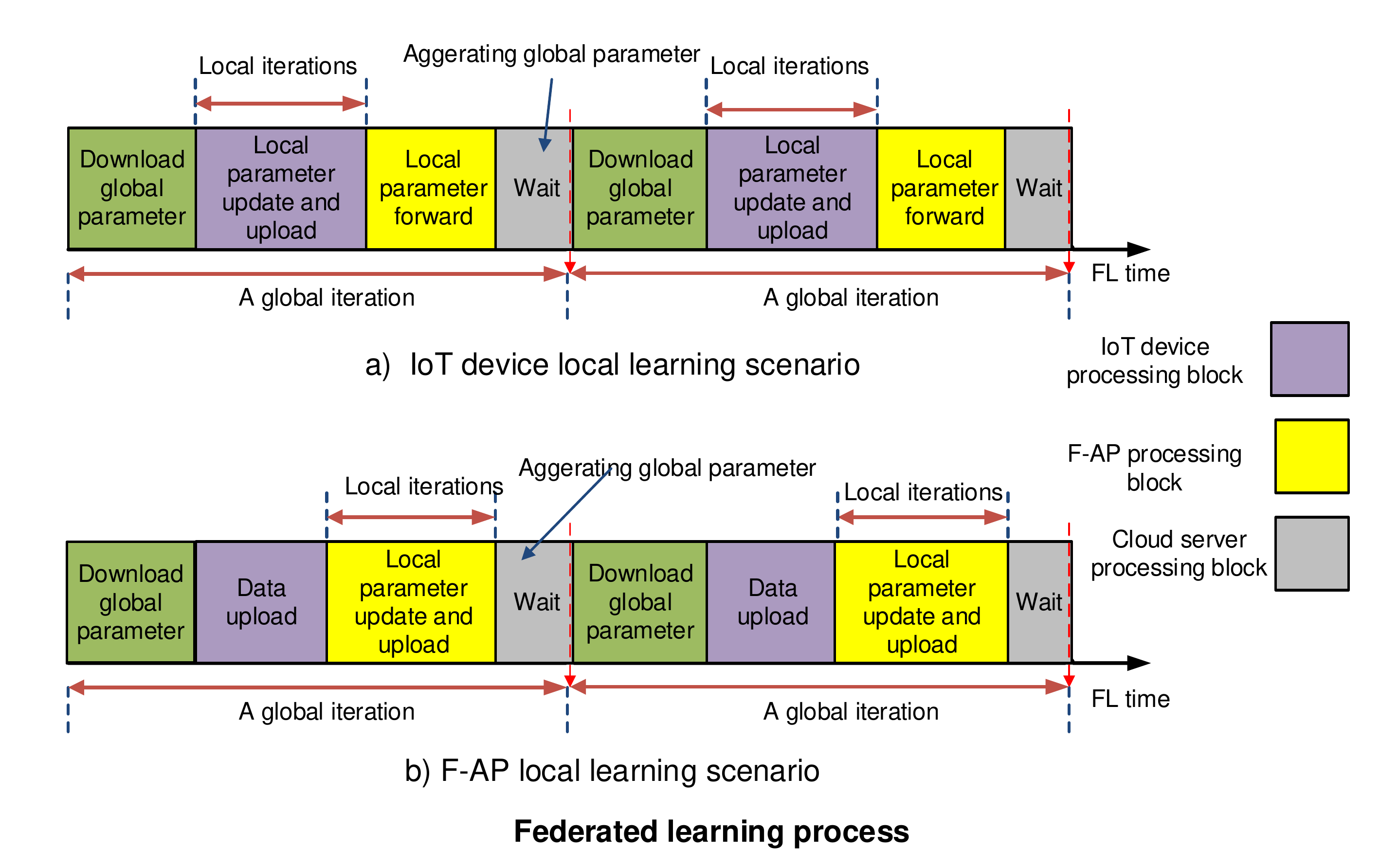}}
		\label{NCT}
	\end{subfigure}
	
	\caption{Illustration of integrated 	FCC-enabled IoT networks and FL process.}
	\label{fig1}
\end{figure}}

In this work, we consider two different scenarios for training the local models: i) IoT device local learning and ii)  F-AP local learning  that are explained  as follows. 

\vspace*{-0.2cm}
\subsection{IoT Device Local Learning Scenario}
In this scenario, the role of IoT devices is to perform local training  and upload the local parameters to the F-APs and to the CS. The F-APs is only responsible for forwarding the collected local parameters from the IoT devices to the CS. This scenario is illustrated in Fig. \ref{fig2a}-(a) and divided into FL time and energy consumption as follows.

\textit{1) FL time:} In each iteration, the FL time consists of\ignore{the aggregation time for aggregating local data at IoT devices,} the computation time for local model training and the
transmission time for uploading local model parameters to the F-APs as well as to the CS. In the local training process, IoT device $n$ trains the local model and updates its local parameter until a local accuracy $\epsilon_l$ is achieved \cite{EE_FL_2}. 
Let $C_n$ denote the number of CPU cycles to process one data sample of IoT device $n$, and accordingly,
the number of CPU cycles required for one local iteration over all data samples is
$C_nD_n$. Therefore, the computation time for one local iteration in
IoT device $n$ can be calculated as $\frac{C_n D_n}{f_n}$, where $f_n$ is the computational speed of the CPU in IoT device $n$ (in cycles per second) \cite{10_new}. Let $\textbf f_\text{N}$ be a $1\times N$ matrix
containing the computation frequency allocations of all IoT devices, i.e., 
$\textbf f_N = [f_n]$. The number of local iterations to reach the local accuracy $\epsilon_l$ in IoT device $n$ is $T_l=\frac{2}{(2-\delta \beta)\delta \vartheta}\ln(1/\epsilon_l)$, where $\delta, \vartheta, \beta$ are constant parameters \cite{7a}. Then, the computation time of IoT device $n$ is expressed as
\begin{equation}
T^c_n=T_l\frac{C_nD_n}{f_n}.
\end{equation}
After performing the local learning at IoT device $n$, suppose that the data size $d_n$ of each
resulting local parameter $\omega_n$ is fixed over the learning process \cite{13}.
Hence, the transmission time of IoT device $n$ for uploading
its parameters to F-AP $k$ on RRB $z$ is
$T^w_n=\frac{d_n}{R_{k,z}^n}$.

Note that the global model parameters can only be updated by the CS
after all local model parameters are received from the F-APs. Consequently, the FL time $\tau_1$ in each global iteration is  determined
by the longest duration time of receiving the parameters among
all IoT devices and the longest duration time of forwarding the parameters from the F-APs to the CS. Moreover, the transmission duration of F-AP $k$ to upload its collected parameters to the CS is $T^w_k=\frac{\sum_{n\in \mathcal N_k}d_n}{R_{fh}}$.
Hence, the learning time $\tau_1$ of one global iteration can be calculated as
\begin{align}
\label{FL_1}
\tau_1&=\max_{n \in \mathcal N}\{T^c_n+T^w_n\}+\max_{k \in \mathcal K}\{T^w_k\}\nonumber \\& =\max_{n \in \mathcal N}\left\{T_l\frac{C_n D_n}{f_n}+\frac{d_n}{R_{k,z}^n}\right\}+\max_{k \in \mathcal K}\left\{\frac{\sum_{n\in \mathcal N_k}d_n}{R_{fh}}\right\}.
\end{align}
\ignore{and, the total FL time of all global iterations $T_g$ is $\tau_1=T_g \tau^l$.}

The learning time $\tau_1$ should satisfy the QoS requirement. Specifically, $\tau_1$ should be no more than the maximum FL time $T_q$, i.e., $\tau_1\leq T_q$. Hence, the QoS requirement can be expressed as
\begin{align}
\max_{n \in \mathcal N}\left\{T_l\frac{C_nD_n}{f_n}+\frac{d_n}{R_{k,z}^n}\right\}+\max_{k \in \mathcal K}\left\{\frac{\sum_{n\in \mathcal N_k}d_n}{R_{fh}}\right\}\leq T_q,
\end{align}
and can be written as
\begin{align}
T_l\frac{C_n D_n}{f_n}+\frac{d_n}{R_{k,z}^n}+\frac{\sum_{n\in \mathcal N_k}d_n}{R_{fh}}\leq T_q, \forall n\in \mathcal N, \forall k\in \mathcal K.
\end{align}

\textit{2) Energy consumption model:}
The IoT device's energy is consumed for both  local model
training and parameter transmission over wireless links that is explained as follows. 
\begin{itemize}

	\item \textit{Local computation:} We adopt the widely used energy
	consumption model which considers that the energy
	consumption of IoT device $n$ to process a single CPU cycle is $\alpha f^2_n$, where $\alpha$	is a constant related to the switched capacitance \cite{14,15}. Hence, the energy consumption of IoT device $n$
	for local computation is $
	E^c_n=T_lC_nD_n\alpha f^2_n$ \cite{7a}.
	\item \textit{Parameter transmission:} The energy consumption to upload local model
	parameters to the F-APs over wireless links can be denoted by $E^w_n$ and calculated as $p_nT^w_n$. Since the local parameters  are forwarded from the F-APs to the CS over high transmission links, the energy consumption is negligible. Hence, we discard the F-APs' energy consumption.
	
\end{itemize}

By combining all the aforementioned terms of energy consumption, the total energy consumption of the system in the IoT device local learning scenario can be	calculated as 
\begin{align}
 E=\sum_{n\in\mathcal N}\left(E^w_n+E^c_n\right)= \sum_{n\in\mathcal N}\left[ \frac{p_nd_n}{R^n_{k,z}}+T_lC_nD_n\alpha f^2_n \right].
\end{align}

\ignore{At the F-APs, the energy consumption of F-AP $k$
	for computation is $E^c_k=T\sum_{n\in \mathcal K_k}(C_nD_n)\alpha f^2_k$ and the energy consumption for uploading local model
	parameters is 
	\begin{equation}
	E^u_k=\frac{p_k\sum_{n\in \mathcal K_k}s_n}{R_k}=\frac{p_k\sum_{n\in \mathcal K_k}s_n}{B\log_2 \left(1+\frac{p_sG^k}{N_0B}\right)},
	\end{equation}
	where $G^k$ denotes the wireless
	channel gain between F-AP $k$ and the fog server. Consequently, the energy consumption of all F-APs can be
	calculated as  $E_{F-APs}=\sum_{k\in\mathcal K}\left(E^c_k+E^u_k\right)$.}

\subsection{F-AP Local Learning Scenario}
The IoT devices, in this scenario, are solely responsible for uploading their data to the F-APs. The F-APs, then, train local models using these collected data and upload the resulting local parameters to the CS for global aggregation. This scenario is illustrated in Fig. \ref{fig2a}-(b). The FL time and energy consumption of this scenario are explained as follows.

\textit{1) FL time:} In each iteration, the FL time consists of the transmission time for uploading the data from IoT devices to F-APs, the computation time for training local models at the F-APs, and the transmission time for uploading the resulting local model parameters to the CS. Consider that the data size $\mathtt D_n$ of the
uploaded data $\mathcal D_n$ by IoT device $n \in \mathcal N_k$ is fixed over the learning process \cite{13}. The transmission time for uploading
 data from IoT devices $\mathcal N_k$ to F-AP $k$ is written as 
\begin{equation}
T^w_k=\max_{n\in \mathcal N_k}\left\{\frac{\mathtt D_n}{R_{k,z}^n}\right\}, \forall z\in \mathcal Z.
\end{equation}
Each F-AP $k$ iteratively trains the local learning model on the collected data and updates its local parameter $\omega_k(t)$ until a local accuracy $\epsilon_l$ is achieved. 
Let $C_k$ denote the number of CPU cycles to process one data sample
of F-AP $k$. Hence,
the number of CPU cycles required for one local iteration is
$C_kB_k$, where $B_k$ is the number of uploaded data samples $\mathcal B_k$ to F-AP $k$, i.e., $B_k=|\mathcal B_k|$ and $\mathcal B_k=\cup_{n\in \mathcal N_k}\mathcal D_n$. Therefore, the computation time for one local iteration at F-AP $k$ can be calculated as $\frac{C_k B_k}{f_k}$, where $f_k$ is the computational speed of the CPU in F-AP $k$ (in cycles per second). Let $\textbf f_\text{K}$ be a $1\times K$ matrix
containing the computation frequency allocations of all F-APs, i.e., 
$\textbf f_K = [f_k]$. Consider that the number of local iterations of F-APs to reach the local accuracy $\epsilon_l$ is $T_l$. Then, the computation time of F-AP $k$ is expressed as
\begin{equation}
T^c_k=T_l\frac{C_kB_k}{f_k}.
\end{equation}

Since the global model parameters can only be updated
after all local model parameters are received from the F-APs,  the FL time $\tau_2$ in each global iteration is determined
by the longest time for uploading data to F-APs, training local models at the F-APs,  and the longest time for transmitting the parameters from the F-APs to the CS. Thus, the learning time $\tau_2$ of one global iteration can be calculated as
\begin{align}
\label{FL_2}
\tau_2&=\max_{k \in \mathcal K}\{T_k^w+T_k^c+T_k\}\nonumber \\&=\max_{k \in \mathcal K}\left\{\max_{n\in \mathcal N_k}\left\{\frac{\mathtt D_n}{R_{k,z}^n}\right\}+T_l\frac{C_k B_k}{f_k}+\frac{d_k}{R_{fh}}\right\},
\end{align}
where $d_k$ is the local parameter size of F-AP $k$. Similar to the first scenario, the learning time $\tau_2$ should satisfy the QoS requirement, i.e., $\tau_2\leq T_q$. Thus, the QoS requirement can be expressed as
\begin{align}
\max_{k \in \mathcal K}\left\{\max_{n\in \mathcal N_k}\left\{\frac{\mathtt D_n}{R_{k,z}^n}\right\}+T_l\frac{C_k B_k}{f_k}+\frac{d_k}{R_{fh}}\right\}\leq T_q,
\end{align}
and can be written per F-AP as
\begin{align}
\max_{n\in \mathcal N_k}\left\{\frac{\mathtt D_n}{R_{k,z}^n}\right\}+T_l\frac{C_k B_k}{f_k}+\frac{d_k}{R_{fh}}\leq T_{q}, \forall  k\in \mathcal K.
\end{align}

\textit{2) Energy consumption model:}
The IoT device's energy is consumed for data transmission over wireless links, which is $E_n^w=p_nT_n^w=\frac{p_n\mathtt D_n}{R_{k,z}^n}$. For the F-APs, the energy consumption is explained as follows.
\begin{itemize}
	\item \textit{Local computation:} The energy
	consumption model of F-AP $k$ for processing a single CPU cycle is $\alpha f^2_k$. Thus, the energy consumption of F-AP $k$
	for local computation is expressed as
	$	E^c_k=T_lC_kB_k\alpha f^2_k$ \cite{7a}.
	\item \textit{Parameter transmission:} The energy consumption for uploading local model
	parameters to the CS is $E^w_k=q_kT^w_k=\frac{q_kd_k}{R_{fh}}$, where $q_k$ is the transmit power of F-AP $k$.
	
\end{itemize}

By combining all the aforementioned terms of energy consumption, the total energy consumption of all IoT devices and F-APs in the second scenario can be	calculated as 
\begin{align}
E&=\sum_{n\in\mathcal N}E^w_n+\sum_{k\in \mathcal K}(E^w_k+E^c_k)\nonumber \\&=\sum_{n\in\mathcal N}\left[ \frac{p_n\mathtt D_n}{R^n_{k,z}}\right]+\sum_{k\in\mathcal K}\left[ T_lC_kB_k\alpha f^2_k +\frac{q_kd_k}{R_{fh}}\right].
\end{align}

\section{Problem Formulation}\label{PF}
\ignore{The role of each IoT device depends on its connectivity to
the F-APs, and preference of the decision-maker to
minimize the consumed
energy. Consequently, the active IoT devices set should be intelligently selected and a suitable role should be assigned to each IoT active device.}

We propose to minimize the total energy consumption for a delay-constrained FL. Specifically, our proposed framework  intelligently selects the active IoT devices that perform local learning and assigns active IoT devices to the suitable F-APs.  
Considering IoT device local learning scenario, the energy  minimization problem  can be  formulated as
\begin{subequations}
	\begin{align} \nonumber 
	& \mathcal{P}_1: 
	\min_{\substack{\mathbf A, \mathbf S, \textbf{f}_N,\mathbf p}}  \sum_{n\in\mathcal N}\left[ \frac{p_nd_n}{R^n_{k,z}}+T_lC_nD_n\alpha f^2_n \right]\\
	&\rm s.t.
	\begin{cases}  \nonumber
	\hspace{0.2cm} \text{C1:}\hspace{0.2cm} \sum_{k\in \mathcal K}a_{k,n} =1 ~\&~ \sum_{z\in \mathcal Z}s^n_{k,z} =1, \forall n \in \mathcal N, \\ 
	\hspace{0.2cm} \text{C2:}\hspace{0.2cm} \sum_{n\in \mathcal N}s^n_{k,z} \leq 2, \forall k\in \mathcal K, z \in \mathcal Z;
	\hspace{0.2cm}\\ \hspace{0.2cm} \text{C3:}\hspace{0.2cm} f^{\min}_n\leq f_n \leq f^{\max}_n, ~\forall n\in \mathcal{N},\\
	\hspace{0.2cm} \text{C4:}\hspace{0.2cm} \tau_1\leq T_q;\\
		\hspace{0.2cm} \text{C5:}\hspace{0.2cm} 0 \leq p_n \leq p_{\max}, ~\forall n\in \mathcal{N};	\hspace{0.2cm}\\ 
	\hspace{0.2cm} \text{C6:}\hspace{0.2cm} a_{i,j}\in \{0,1\}, s^k_{i,j}\in \{0,1\}.
	\end{cases}
	\end{align}
\end{subequations}
In $\mathcal P_1$, C1 indicates that each IoT device is scheduled to only one F-AP and to only one RRB in that F-AP; C2 indicates that maximum two IoT devices can be scheduled to each F-AP   at the same time; C3 is the constraint on local computation resource allocations of IoT devices; C4 indicates the QoS requirement on the FL time; and C5 is the transmit power control constraint. 

The optimization problem of energy consumption minimization for FL integrated
FCC-enabled IoT networks of the F-AP local learning scenario can be expressed as
\begin{subequations}
	\begin{align} \nonumber 
	& \mathcal{P}_2: 
	\min_{\substack{\mathbf A, \mathbf S, \textbf{f}_K,\bf p}} \sum_{n\in\mathcal N}\left[ \frac{p_n\mathtt D_n}{R^n_{k,z}}\right]+\sum_{k\in\mathcal K}\left[ T_lC_kB_k\alpha f^2_k +\frac{q_kd_k}{R_{fh}}\right]\\
	&\rm s.t.
	\begin{cases}  \nonumber
	\hspace{0.2cm} \text{C1}, \text{C2}, \text{C5}, \text{C6},\\
	\hspace{0.2cm} \text{C3:}\hspace{0.2cm} f^{\min}_k \leq f_k \leq f^{\max}_k, ~\forall k\in \mathcal{K},\\
	\hspace{0.2cm} \text{C4:}\hspace{0.2cm} \mathtt N_k \leq U, ~\forall k\in \mathcal{K};\\
	\hspace{0.2cm} \text{C7:}\hspace{0.2cm}\tau_2\leq T_q.
	\end{cases}
	\end{align}
\end{subequations}
In $\mathcal P_2$, C3 is the constraint on local computation resource allocations of F-APs; C4 represents that the number of scheduled IoT devices to F-AP $k$ $\mathtt N_k$ is less than or equal to the maximum number of scheduled IoT devices $U$. This is becasue each F-AP has certain computation frequency capability, and thus it can schedule only a limited number of IoT devices.  Finally, C5 indicates the QoS requirement on the FL time. Note that both problems $\mathcal P_1$ and $\mathcal P_2$ are non-convex optimization problems. In addition, owing to the coupling of  the optimization variables $f_n$, $f_k$ and $p_n$, it is challenging to solve  problems $\mathcal P_1$ and $\mathcal P_2$.  To this end, we divide both optimization problems into two subproblems and optimize them iteratively in order to achieve suboptimal yet practical solutions.

\section{Energy Consumption Minimization: First Scenario}\label{G}
\subsection{Problem $\mathcal P_1$ Transformation}\label{G-A}
Solving problem $\mathcal P_1$ owing to its mixed combinatorial characteristics  is challenging. Although  exhaustive
search and branch-and-bound approaches can obtain near-optimal solution to $\mathcal P_1$, such approaches are not suitable for the practical systems due to  the significantly increased  computational complexity. To strike a suitable balance between the required complexity and performance, we propose an iterative approach to solve problem $\text{P1}$  for large-scale IoT networks. To this end, we decompose $\mathcal P_1$ into the following two subproblems, namely, (i) IoT device scheduling and power allocation subproblem for a given IoT device' computation frequency allocation, and (ii) IoT device' computation frequency allocation subproblem for the determined power and IoT device scheduling.


\textbf{IoT Device Scheduling and Power Allocation Subproblem:} For a fixed set of computation frequency allocation, $ f^*_n, \forall n\in \mathcal N$,	the optimization problem $\mathcal P_1$ can be written as
\begin{subequations}
	\begin{align} \nonumber 
	& \mathcal{P}_3: 
	\min_{\substack{\mathbf A, \mathbf S,\bf p}} \sum_{n\in\mathcal N} \frac{p_nd_n}{W\log_{2}(1+\gamma^n_{k,z})} \\
	&\rm s.t.
	\begin{cases}  \nonumber
	\hspace{0.2cm} \text{C1}, \text{C2}, \text{C5},\\
	\hspace{0.2cm} \text{C4:}\hspace{0.2cm} T_l\frac{C_n D_n}{f^*_n}+\frac{d_n}{R_{k,z}^n}\leq T_{q,k}, \forall n\in \mathcal N,
	\end{cases}
	\end{align}
\end{subequations}
where $T_{q,k}=T_q- \frac{\sum_{n\in \mathcal N_k}d_n}{R_{fh}}$. In $\mathcal{P}_3$, the optimization is over the continuous variables $\bf p$, and the discrete variables $a_{k,n}$, and $s^n_{k,z}, \forall k\in \mathcal K, n\in \mathcal N, z\in \mathcal Z$.  Nevertheless, it is still challenging to solve problem $\mathcal P_3$
because of the non-convexity.  We hence design an efficient yet low-complexity graph theory algorithm to tackle this problem  in Section IV. B.

\textbf{Computation Frequency Allocation Subproblem:}
For the given transmit power allocation and scheduling among the IoT devices, RRBs, and F-APs, problem $\mathcal P_1$  is reduced to the following subproblem
\begin{subequations}
	\begin{align} \nonumber 
	& \mathcal{P}_4: 
	\min_{\substack{\textbf{f}_N}} \sum_{n\in\mathcal N} T_lC_nD_n\alpha f^2_n \\
	&\rm s.t.
	\begin{cases}  \nonumber
		\hspace{0.2cm} \text{C3:}\hspace{0.2cm} f^{\min}_n \leq f_n \leq f^{\max}_n, ~\forall n\in \mathcal{N},\\
		\hspace{0.2cm} \text{C4:}\hspace{0.2cm} \max_{n \in \mathcal N}\left\{T_l\frac{C_n D_n}{f_n}+\frac{d_n}{R_{k,z}^{*n}}\right\}\leq T_{q,k}.
	\end{cases}
	\end{align}
\end{subequations}
C4 can be transformed into  $T_l\frac{C_n D_n}{f_n}+\frac{d_n}{R_{k,z}^{*n}}\leq T_{q,k}, \forall n \in \mathcal N$. Hence, the
lower bound of IoT device's computation frequency can
be calculated as $f_n\geq \frac{T_lC_n D_n}{T_{q,k}-\frac{d_n}{R_{k,z}^{*n}}}$. For simplicity, we denote $\hat f_n = \frac{T_lC_n D_n}{T_{q,k}-\frac{d_n}{R_{k,z}^{*n}}}$. Then, $f_n$ satisfies  $f_n\geq \max\left\{f^{\min}_n,\hat f_n\right\}$, and accordingly, C3 and C4 can be combined as $\max\left\{f^{\min}_n,\hat f_n\right\}\leq f_n\leq f^{\max}_n$. Therefore, $\mathcal{P}_4$ can be expressed as
\begin{subequations}
	\begin{align} \nonumber 
	& \mathcal{P}_5: 
	\min_{\substack{f_n}} \sum_{n\in\mathcal N} T_lC_nD_n\alpha f^2_n\\
	&\rm s.t.
	\hspace{0.2cm} \max\left\{f^{\min}_n,\hat f_n\right\} \leq f_n \leq f^{\max}_n, ~\forall n\in \mathcal{N}.
	\end{align}
\end{subequations}


\textit{\textbf{Lemma 1:}}
\textit{The closed-form solution of subproblem $\mathcal P_5$ is obtained as}
\begin{equation}
\label{close_form_P5}
\begin{split}
f_n=\begin{cases}
& f^{\min}_n,  ~\text{if} ~\hat f_n \leq f^{\min}_n \\
& \hat f_n , ~\text{if} ~ f^{\min}_n< \hat f_n <  f^{\max}_n\\
&  f^{\max}_n, ~\text{if} ~ \hat f_n  \geq  f^{\max}_n
\end{cases}
\end{split}
\end{equation}	

\proof The proof is omitted due to the space limitation.


%

 $\mathcal P_1$ is solved by iteratively solving both subproblems $\mathcal P_3$ and $\mathcal P_5$ until convergence. The overall algorithm to obtain a suitable solution to problem  $\mathcal P_1$ is provided in \sref{G-C}.

\vspace*{-0.4cm}
\subsection{Subproblem $\mathcal P_3$ Solution }\label{G-B}
In this subsection, we develop an effective and low complexity  approach to solve the joint IoT device scheduling and power control subproblem $\mathcal P_3$. Our developed solution designs a graph for all  IoT device-RRB-F-AP feasible schedules and efficiently allocates power levels for the IoT devices in each schedule.  The two-stage solution is explained as follows.
 
\textbf{Stage 1: IoT Device Feasible Scheduling:} In this stage, we design a graph that judiciously 
generates feasible NOMA clusters and jointly optimizes IoT devices-F-APs/RRBs assignments and transmit power of the IoT devices. This stage consists of graph design and maximum weight independent search (MWIS) method.

\textit{1) \textbf{Graph design:}} Let $\mathcal G=(\mathcal V, \mathcal E, \mathcal W)$ represents an undirected graph. The graph $\mathcal G$ is constructed by generating a vertex $v$ for each 2-IoT devices, RRB, and F-AP in the network as follows. We start from RRB $z=1$, and assume that IoT device $n=1$ is allocated to it. Then, we find the available NOMA clusters according to the possible two scenarios:
\begin{enumerate}
	\item  If IoT device $n=1$ is not in the service area of the $k$-th F-AP, we check IoT device $n=2$ for possible association to RRB $z$ and F-AP $k$, and then continue finding the second IoT device.
	
	\item  If IoT device $n=1$ is in the service area of the $k$-th F-AP, then we find the second UD $j=n+1$ (currently, $j = 2$), for the ($n = 1, z = 1$) pair. Afterwords, we find $p^*_n$, $p^*_j$ and calculate the rates, and then, we generate a vertex $v=\{(r^*_n, z,k), (r^*_j, z,k)\}$ that represents a feasible NOMA cluster. Given $r^*_n$, $r^*_j$, we then compute the weight of that vertex $w(v)= X_{n}+X_{j}$ and update $\mathcal G$. If adding $j=2$ is infeasible, we let
	$j=j+1=3$, and  we verify the 	feasibility and repeat the aforementioned step.

\end{enumerate}
In order to obtain all the feasible NOMA clusters $((n,j)\in \mathcal N, z\in \mathcal Z, k \in \mathcal K), j > n$, we iteratively repeat the above process (1), (2). The vertices in the designed grpah $\mathcal G$ that represent NOMA feasible clusters are connected by a conflict edge according to the following connectivity conditions (CCs):
\begin{itemize}
	\item \textbf{CC1:} The same IoT devices (any IoT device or both IoT devices) are associated with both vertices $v$ and $v'$.
	\item \textbf{CC2:} The same RRB in the same F-AP is associated with both vertices $v$ and $v'$.
\end{itemize} 
In summary, two distinct vertices $v$ and $v'$ representing two different NOMA clusters are connecting by a conflict edge if and only if the associations of RRBs and IoT devices (one or both associations) they represent are appeared in both vertices.

To select the IoT device-RRB-F-AP scheduling that provides a local minimum energy consumption, we design a proper weight
$w(v)$ to each vertex $v \in \mathcal{G}$.  For notation simplicity, we define the utility
of IoT devices $n$ and $j$ as $X_n=T_lC_nD_n\alpha f^2_n+  \frac{p_nd_n}{R^n_{k,z}}$, $X_j=T_lC_jD_j\alpha f^2_j+  \frac{p_jd_j}{R^j_{k,z}}$, respectively. Therefore, the weight of vertex $v$ that reflects
the minimum energy consumption of IoT devices $n, j$ can be given by 
\begin{equation}\label{W(v)}
w(v)= X_{n^v}+X_{j^v},
\end{equation}
where $X_{n^v}$ and $X_{j^v}$ are the utility of associated IoT devices $n$ and $j$ to vertex $v$, respectively. The weight of vertex $v$ in \eref{W(v)} is determined by the transmit powers $\{p^*_{n}$, $p^*_{j}\}$, computation frequency allocation $\{f^*_n$, $f^*_j\}$, RRB $z^v$, and F-AP $k^v$ allocated to them.

\textit{2) \textbf{MWIS search method:}} In this step, the algorithm itratively and greedily selects the MWIS $\Gamma^*$ among all the minimal independent sets $\Gamma$ in the graph $\mathcal G$, where in each iteration we implement the following procedures. We compute the weight of all generated vertices using \eref{W(v)}. The vertex with the minimum weight $v^*$ is selected among all other corresponding vertices. The selected vertex $v^{*}$ is, then, added to $\Gamma^*$ that is initially empty. Afterwards, we update the $\mathcal{G}$ graph by removing the selected vertices $v^{*}$ and its connected vertices. As such, the next selected vertex is not in conflict connection with the already selected vertices in $\Gamma^*$. The process continues until no more vertices exist in $\mathcal{G}$. Since each RRB in each F-AP contributes by a single vertex, the number of vertices in $\Gamma^*$ is $ZK$.

\textbf{Stage 2: Power Allocation:} From the designed $\mathcal G$, we obtain a set of vertices that represent NOMA clusters.  Each NOMA cluster includes two IoT devices that simultaneously transmit to an F-AP over an RRB.  For each vertex, we aim to determine transmit power allocations of the IoT devices such that (i) the overall  uplink transmission rate is improved by suppressing the interference between the IoT devices and (ii) the energy consumption for wireless transmission is reduced. Without loss of generality, we consider a vertex where the $n$-th and the $j$-th IoT devices are clustered, and both IoT devices transmit to the $k$-th F-AP over the $z$-th RRB. For such a vertex, we formulate the  transmit power allocation subproblem as $\mathcal{P}_3-1$ at the top of the next page.
\begin{table*}
	\vspace*{-0.4cm}
	\begin{normalsize} 
\begin{equation}
\label{P_3_2}
\begin{split} 
\mathcal{P}_3-1: &
\max_{\substack{0 \leq p_n\leq p_{\max},\\ 0\leq p_j \leq p_{\max}}}  W\left(\log_{2}(1+\gamma^n_{k,z})+\log_{2}(1+\gamma^j_{k,z})\right)-V\left(p_n+p_j\right)\\
& \text{s.t.}
\begin{cases}
&\text{C8:}~W\log_{2}(1+\gamma^n_{k,z}) \geq R_{th,n}\\
& \text{C9:}~W\log_{2}(1+\gamma^j_{k,z}) \geq R_{th,j}.
\end{cases}
\nonumber
\end{split}
\end{equation}
	\end{normalsize}
\vspace*{-0.2cm}
\hrulefill
\end{table*}
In subproblem $\mathcal{P}_3-1$, $R_{th,n}$ and $R_{th,j}$ are the required uplink data rates for the $n$-th and $j$-th IoT devices, respectively;  and $V$ is a given weight factor. In particular, $R_{th,n}=\frac{T_gd_n}{T_{q,k}-T_gT_l\frac{C_nD_n}{f_n^*}}$ and $R_{th,j}=\frac{T_gd_j}{T_{q,k}-T_gT_l\frac{C_jD_j}{f_j^*}}$. Essentially,  the rate constraints C8 and C9 satisfy the FL delay constraint C4. On the other hand, the weight factor $V$ is selected to strike a suitable balance between capacity and energy consumption of the vertices.

Note that the power allocation depends on the channel gain of the associated IoT devices. To this end, we first define $\Delta_n=\frac{\sigma^2}{p_{\max}}\left(2^{R_{th,n}/W}-1\right)$ and $\Delta_j=\frac{\sigma^2}{p_{\max}}\left(2^{R_{th,j}/W}-1\right)$. Thereafter, we consider the following four cases: \textbf{Case I:} $\left|G_{k,z}^n\right|^2 <\Delta_n$ and $\left|G_{k,z}^j\right|^2 <\Delta_j$, \textbf{Case II:} $\left|G_{k,z}^n\right|^2  \geq\Delta_n$ and $\left|G_{k,z}^j\right|^2 <\Delta_j$, \textbf{Case III:} $\left|G_{k,z}^n\right|^2 <\Delta_n$ and $\left|G_{k,z}^j\right|^2  \geq\Delta_j$, and \textbf{Case IV:} $\left|G_{k,z}^n\right|^2  \geq \Delta_n$ and $\left|G_{k,z}^j\right|^2  \geq\Delta_j$. The transmit power allocations, $(p_n^*, p_j^*)$, for each case are given as follows

\textbf{Case I:} In this case, both IoT devices can not satisfy the rate constraints even using the maximum transmit power. Consequently, both IoT devices suspend their data transmission, and we obtain $p_n^*=0$ and $p_j^*=0$.

\textbf{Case II:} In this case, only the $n$-th IoT device can satisfy the required rate constraint, and the transmission of the $j$-th IoT device is suspended. Therefore, we obtain $p_j^*=0$ and $p_n^*=\frac{\sigma^2}{|G_{k,z}^{n}|^2}\left(2^{R_{th,j}/W}-1\right)$.

\textbf{Case III:} In this case, only the $j$-th IoT device can satisfy the required rate constraint, and the transmission of the $n$-th IoT device is suspended. Therefore, we obtain $p_n^*=0$ and $p_j^*=\frac{\sigma^2}{|G_{k,z}^{j}|^2}\left(2^{R_{th,j}/W}-1\right)$.

\textbf{Case IV:} In this case, both IoT devices can simultaneously transmit. Without loss of generality, we assume that $|G_{k,z}^{j}|^2< |G_{k,z}^{n}|^2$, i.e., the $n$-th IoT device has a better channel gain compared to the $j$-th IoT device. According to the NOMA principle, the $k$-th F-AP first decodes the $n$-th IoT device's signal, and subsequently, decodes the $j$-th IoT device's signal after removing the interference from the $n$-th device via applying the SIC technique. We first introduce the following lemma to update the  $j$-th IoT device's power allocation 

\textit{\textbf{Lemma 2:}}
\textit{Assume that the given transmit power allocations for the $n$-th and the $j$-th IoT devices are $\tilde{p}_n$ and $\tilde{p}_j$, respectively. Therefore, the $j$-th IoT device's transmit power allocation to maximize subproblem $\mathcal P_3-1$ is obtained as
	\begin{equation}
	\label{p_j}
	p_j=\left[\frac{\frac{\gamma_{k,z}^{j}}{1+\gamma_{k,z}^{j}}}{V+\frac{(\gamma_{k,z}^{n})^2}{1+\gamma_{k,z}^{n}}\frac{|G_{k,z}^{j}|^2}{\tilde{p}_n |G_{k,z}^{n}|^2}}\right]_{p_{th}}^{p_{\max}}
	\end{equation}
	where $\gamma_{k,z}^{n}$ and $\gamma_{k,z}^{j}$ are calculated by plugging   $\tilde{p}_n$ and $\tilde{p}_j$ to \eqref{SNR}, $p_{th}=\frac{\sigma^2}{|G_{k,z}^{j}|^2}\left(2^{R_{th,j}/W}-1\right)$, and $[\cdot]_{p_{th}}^{p_{\max}}$ denotes projection in the range of $[p_{th}, p_{\max}]$.}

\proof The proof \ignore{is similar to  \cite[Lemma 2]{Ahmed_Multi_level_EC}, and }is omitted due to the space limitation.

We consider a suboptimal approach to iteratively update the transmit power allocation of both the $n$-th and the $j$-th IoT devices in inner and outer loop. Specifically, using a bi-section search method, the outer loop adjusts the power allocation of the $n$-th IoT device  such that the rate constraint $\text{C8}$ is satisfied, and the inner loop adjusts the power allocation of the $j$-th IoT device according to \textit{Lemma 2}.   Let us denote the  minimum and maximum power level for the $n$-th IoT device as $p_{n,low}$ and $p_{n,high}$. The initial transmit power of the $n$-th IoT device is obtained as $p_n=\frac{p_{n,low}+p_{n,high}}{2}$. By plugging the transmit power of the $n$-th IoT device to \textit{Lemma 2}, the $j$-th IoT device's transmit power is determined. Thereafter, the achievable rate of the $n$-th IoT device is calculated. If $W \log_{2}(1+\gamma_{k,z}^{n})>R_{th,1}$, $p_{n,high} \leftarrow p_n$ is applied and if $W \log_{2}(1+\gamma_{k,z}^{n})<R_{th,1}$, $p_{n,low} \leftarrow p_n$ is applied. Then, the  transmit power of the $n$-th IoT device is updated as $p_n=\frac{p_{n,low}+p_{n,high}}{2}$. The aforementioned procedures are repeated until $\left|W \log_{2}(1+\gamma_{k,z}^{n})-R_{th,1}\right|$ approaches a small value. The final values of $p_n$ and $p_j$ provide the required power allocations for the Case IV. The overall two-stage algorithm to the subproblem $\mathcal P_3$ is summarized in Algorithm \ref{alg1}.

\begin{algorithm}[t!]

	\textbf{Data:} $\mathcal{N}, \mathcal{K}, \mathcal{Z}$, $G^n_{k,z}$, $p_\text{max}$, and $f^*_n$,  $(n,k,z)\in\mathcal{N}\times\mathcal{K}\times\mathcal{Z}$.\;
	 \textbf{Stage 1: IoT device feasible scheduling}\;
		\begin{itemize}
		\item Initialize $\mathcal G=\emptyset$. \\
		\For{$k=1:K$}{
		
		\For{$z=1:Z$}{
	Set $n=1$\\
		\eIf{the $n$-th IoT device in $\mathcal N_k$}{
		 Set $j=n+1$\\
		\While{$j<N$}{
		\If{the $j$-th IoT device in $\mathcal N_k$}{
		Based on $p_n$ and $p_j$, calculate $r_n, r_j$\\
		Generate vertex $v=\{(r^*_n, z,k), (r^*_j, z,k)\}$ and set $\mathcal G\longleftarrow \mathcal G\cup v$
	}
		$j=j+1$
	    }}
   {$n=n+1$}

	}
	}
		\item  For each $v \in \mathcal V$, finds its neighborhood $\mathcal N_{\mathcal G}(v)$ according to \textbf{CC1}  and \textbf{CC2}.
		\item Calculate the weight of each vertex $w(v)$ as in \eref{W(v)}.
		\item Let $\Gamma^* = \emptyset, l=0, \mathcal G_l=\mathcal G$.\\
		\item \textbf{MWIS Search Method:}\\
		\While{$\mathcal V(\mathcal G_l)\neq \emptyset$}{
	 $v^*=\arg\min_{{v\in \mathcal G_l}(\Gamma)} \{w (v)\}$ and set $\Gamma \leftarrow \Gamma \cup v^*$
		
		 Let $\mathcal V(\mathcal G_{l+1})=\mathcal V(\mathcal G_l(\Gamma))$\\
		 $l=l+1$
	}
		\item Output: The MWIS and its corresponding IoT device scheduling.
		\end{itemize}
	 \textbf{Stage 2: Power allocation:} Allocate transmit power in the NOMA clusters according to the method described in \sref{G-B}.

	\caption{Low Complexity Graph Algorithm} \label{alg1}
\end{algorithm}
\begin{algorithm}[t!]
	\begin{algorithmic}[1]
		\STATE \textbf{Input:} $\mathcal{N}, \mathcal{K}, \mathcal{Z}$, $C_nD_n$, $T_l$, $W, \sigma^2$, $p_{\max}, T_q$, $f^{\min}_n$, and $f^{\max}_n$.\;
		\STATE \textbf{Output:} IoT device scheduling, $p_n$, and $f_n$.\;
		\STATE Initialize the number of iteration $t=1$, $f^{(0)}_n=f^{\min}_n$,  $f^{(1)}_n=f^{\max}_n, \forall n\in \mathcal N$.\;
		\WHILE{$f^{(t)}_n \neq f^{(t-1)}_n ~and~ t< T_{\max}$}
		\STATE Solve $\mathcal P_3$ as in Algorithm \ref{alg1}.\;
		
		\STATE Calculate the solution $f^{(t)}_n$ of the 
		problem $\mathcal P_5$ according to \textit{Lemma 1}.\; 
		\STATE $t=t+1$.
		\ENDWHILE
		\STATE Return IoT device scheduling, $p^{(t)}_n$, and $f^{(t)}_n$. \;
	\end{algorithmic}
	\caption{Proposed Iterative Algorithm} \label{alg2}
\end{algorithm}

\vspace*{-0.5cm}
\subsection{Proposed Algorithm}\label{G-C}
Our proposed iterative scenario to the problem $\mathcal P_1$ is summarized in
Algorithm \ref{alg2}. Line 4 initiates the computation frequency allocation. The loop in lines 5-9 alternatively obtains the solutions of
subproblems $\mathcal P_3$ and $\mathcal P_5$ and terminates when $f^{(t)}_n$ does not change or the maximum number of iteration is reached. Specifically, line $6$ calculates
$p^{(t)}_n$ with fixed $f^{(t-1)}_n$ from the previous iteration and line $7$ calculates $f^{(t)}_n$ with fixed $p^{(t)}_n$
from the current iteration.
\vspace*{-0.4cm}
\subsection{Complexity Analysis}
The computational complexity of Algorithm \ref{alg2} is dominated by the required complexity of the graph construction stage of Algorithm \ref{alg1}. In order to generate all the vertices using the low-complexity graph punning method, a total of $\mathcal{O}(NKZ)$ computational complexity is required.  The required complexity of connecting the generated vertices, i.e., the required complexity of finding neighborhood  of the generated vertices is $\mathcal{O}((NKZ)^2)$. Therefore, the overall computational complexity of Algorithm \ref{alg2} is obtained as $\mathcal{O}\left(NKZ+(NKZ)^2\right)\approx \mathcal{O}(N^2K^2 Z^2)$.  

\section{Energy Consumption Minimization: Second Scenario}\label{MT}
\subsection{Problem $\mathcal P_2$ Transformation}\label{MT-A}
W decompose $\mathcal P_2$ into two subproblems, namely, (i) joint power allocation and IoT device-F-AP/RRB scheduling optimization subproblem for fixed F-APs' computation frequency allocation, and (ii) F-APs' computation frequency allocation that optimizes computation allocation. 

\textbf{IoT Device Scheduling and Power Allocation Subproblem:} For a fixed set of F-APs' computation frequency allocation, $f^*_k, \forall k\in \mathcal K$,	the optimization problem $\mathcal P_2$ can be written as
\begin{subequations}
	\begin{align} \nonumber 
	& \mathcal{P}_6: 
	\min_{\substack{\mathbf A, \mathbf S,\bf p}} \sum_{n\in\mathcal N} \frac{p_n\mathtt D_n}{W\log_{2}(1+\gamma^n_{k,z})} \\
	&\rm s.t.
	\begin{cases}  \nonumber
	\hspace{0.2cm} \text{C1}, \text{C2}, \text{C4}, \text{C5},\\
	\hspace{0.2cm} \text{C7:}\hspace{0.2cm} \max_{n\in \mathcal N_k}\left\{\frac{\mathtt D_n}{R_{k,z}^n}\right\}+T_l\frac{C_k B_k}{f^*_k}+\frac{d_k}{R_{fh}}\leq T_{q}, \forall  k\in \mathcal K.
	\end{cases}
	\end{align}
\end{subequations}
\ignore{Note that the transmission time for uploading local parameters from F-APs to the CS is ignored in C7 becasue it negligible.} In $\mathcal{P}_6$, the optimization is over the continuous variables $\bf p$, and the discrete variables $a_{k,n}$, and $s^n_{k,z}, \forall k\in \mathcal K, n\in \mathcal N, z\in \mathcal Z$. It is still difficult to solve problem $\mathcal P_6$
because of its non-convexity. To find a tractable solution to $\mathcal P_6$, we develop an efficient algorithm in \sref{MT-B}.

\textbf{Computation Frequency Allocation Subproblem:}
After obtaining $p_n, \forall n \in \mathcal N$ and IoT device-RRB-F-AP scheduling, problem $\mathcal P_2$ is reduced to
\begin{subequations}
	\begin{align} \nonumber 
	& \mathcal{P}_7: 
	\min_{\substack{\textbf{f}_K}} \sum_{k\in\mathcal K}\left[ T_lC_kB_k\alpha f^2_k +\frac{q_kd_k}{R_{fh}}\right] \\
	&\rm s.t.
	\begin{cases}  \nonumber
	\hspace{0.2cm} \text{C3:}\hspace{0.2cm} f^{\min}_k \leq f_k \leq f^{\max}_k, ~\forall k\in \mathcal{K},\\
	\hspace{0.2cm} \text{C7:}\hspace{0.2cm} \max_{k\in \mathcal K}\left\{\max_{n\in \mathcal N_k}\left\{\frac{\mathtt D_n}{R_{k,z}^{*n}}\right\}+T_l\frac{C_k B_k}{f_k}+\frac{d_k}{R_{fh}}\right\}\leq T_{q}.
	\end{cases}
	\end{align}
\end{subequations}
Subproblem $\mathcal{P}_7$ can be equivalently expressed as
\begin{subequations}
	\begin{align} \nonumber 
	& \mathcal{P}_8: 
	\min_{\substack{f_k}} \sum_{k\in\mathcal K}\left[ T_lC_kB_k\alpha f^2_k +\frac{q_kd_k}{R_{fh}}\right]\\
	&\rm s.t.
\hspace{0.2cm} \max\left\{f^{\min}_k,\hat f_k\right\} \leq f_k \leq f^{\max}_k, ~\forall k\in \mathcal{K}, 
	\end{align}
\end{subequations}
where $\hat f_k= \frac{T_lC_k B_k}{T_q-\max_{n\in \mathcal N_k}\left(\frac{\mathtt D_n}{R_{k,z}^{*n}}\right)-\frac{d_k}{R_{fh}}}$.

\textit{\textbf{Lemma 3:}}
\textit{The closed-form solution to $\mathcal P_8$ is obtained as}
\begin{equation}
\label{close_form_P}
\begin{split}
f_k=\begin{cases}
& f^{\min}_k,  ~\text{if} ~\hat f_k \leq f^{\min}_k \\
& \hat f_k , ~\text{if} ~ f^{\min}_k< \hat f_k <  f^{\max}_k\\
&  f^{\max}_k, ~\text{if} ~ \hat f_k  \geq  f^{\max}_k
\end{cases}
\end{split}
\end{equation}	

\proof The proof is omitted due to the space limitation.

%
%
%
\subsection{Subproblem $\mathcal P_6$ Solution}\label{MT-B}
This subsection first addresses the optimization subproblem $\mathcal P_6$ as an IoT device coordinated scheduling problem only, and can be written as
\begin{subequations}
	\begin{align} \nonumber 
	& \mathcal{P}_{9}: 
	\min_{\substack{\mathbf A, \mathbf S}} \sum_{n\in\mathcal N} \frac{p^*_{n}\mathtt D_n}{W\log_{2}(1+\gamma^n_{k,z})} \\
	&\rm s.t.
	\begin{cases}  \nonumber
	\hspace{0.2cm} \text{C1:}\hspace{0.2cm} \sum_{k\in \mathcal K}a_{k,n} =1 ~\&~ \sum_{z\in \mathcal Z}s^n_{k,z} =1, \forall n \in \mathcal N, \\ 
	\hspace{0.2cm} \text{C2:}\hspace{0.2cm} \sum_{n\in \mathcal N}s^n_{k,z} \leq 2, \forall k\in \mathcal K, z \in \mathcal Z\\
		\hspace{0.2cm} \text{C4:}\hspace{0.2cm} \mathtt N_k \leq U, ~\forall k\in \mathcal{K},\\
	\hspace{0.2cm} \text{C7:}\hspace{0.2cm} \max_{n\in \mathcal N_k}\left\{\frac{\mathtt D_n}{R_{k,z}^n}\right\}\leq T_{q,k}, \forall  n\in \mathcal N,
	\end{cases}
	\end{align}
\end{subequations}
where $T_{q,k}=T_q-\left(T_l\frac{C_k B_k}{f^*_k}+\frac{d_k}{R_{fh}}\right)$. The optimization is carried over the variables $\mathbf A$, $\mathbf S$.

On the other hand, for the resulting IoT device-RRB/F-AP
schedule, $\mathcal P_6$ can be considered as a power allocation step and simplifies per
RRB basis. For each RRB $z$, the optimization problem $\mathcal P_6$ can be written as
\begin{subequations}
	\begin{align} \nonumber 
	& \mathcal{P}_{10}: 
	\min_{\substack{\bf p}} \sum_{n\in\mathcal N} \frac{p_n\mathtt D_n}{W\log_{2}(1+\gamma^n_{k,z})} \\
	&\rm s.t.
	\begin{cases}  \nonumber
	\hspace{0.2cm} \text{C5:}\hspace{0.2cm} 0 \leq p_n \leq p_{\max}, ~\forall n\in \mathcal{N},\\ 
		\hspace{0.2cm} \text{C7:}\hspace{0.2cm} \max_{n\in \mathcal N_k}\left\{\frac{\mathtt D_n}{R_{k,z}^n}\right\}\leq T_{q,k}, \forall  n\in \mathcal N,
	\end{cases}
	\end{align}
\end{subequations}
where the optimization is over the set of powers $p_n$, $\forall n\in \mathcal N$. Note that solving the power allocation problem $\mathcal P_{10}$ for the resulting IoT device-RRB/F-AP schedule is omitted in this section becasue it can follow the solution of $\mathcal P_{3}-1$ in  \sref{G-B}.

The graph-based solution of the IoT coordinated scheduling problem $\mathcal P_9$ is explained as follows.

\textit{1) \textbf{IoT Device Coordinated Scheduling:}} The NOMA-coordinated graph is introduced to jointly consider NOMA cluster per RRB, maximum number of IoT devices scheduled to F-AP, and transmission conflict. The NOMA-coordinated graph, denoted by $\mathcal{G}_\text{NOMA}(\mathcal{V},\mathcal{E})$, is designed by
generating all vertices for the $k$-th F-AP.  The vertex set $\mathcal V$ of the entire graph is the union of
vertices of all F-APs.  Consider, for now, generating the vertices of F-AP $k$. Therefore, each vertex $v_{k,z,n,j}$ is generated for each $z\in \mathcal{Z}$ and for every 2-IoT devices $(n,j)$ in the service area of F-AP $k$. Similarly, we generate
all vertices for all F-APs in $\mathcal K$. The configuration of the set of edges in the NOMA-coordinated graph is divided into IoT devices' association and transmission conflict edges. Two vertices $v_{k,z,n,j}$ and $v_{k,z',n^\prime,j\prime}$ representing different RRB $z$ and the same F-AP $k$ are adjacent by a conflict link if the number of scheduled IoT devices $\mathtt N_k$ to F-AP $k$ is more than $U$.  Similarly, two vertices $v_{k,z,n,j}$ and $v_{k^\prime,z^\prime,n^\prime,j^\prime}$ are adjacent by a transmission conflict link if one of these conditions is true:
\begin{itemize}
	\item $n=n'$ and/or $j=j'$. This condition schedules different IoT devices to different RRBs/F-APs.
	\item $z=z'$ and $k = k'$. This condition insists that same RRB in the same F-AP is associated with both vertices $v_{k,z,n,j}$ and $v_{k^\prime,z^\prime,n^\prime,j^\prime}$.
\end{itemize} 
Therefore, two vertices $v_{k,z,n,j}$ and 
$v_{k^\prime,z^\prime,n^\prime,j^\prime}$ are adjacent by
a conflict edge in $\mathcal E$ if they satisfy one of the following \textbf{CCs}.
\begin{itemize}
	\item \textbf{CC1:} ($n \neq n'$ and $j \neq j'$) and ($k = k'$ and $z \neq z'$), we have $\mathtt N_k > U$.
	\item \textbf{CC2:} $n=n'$ and/or $j=j'$.
	\item \textbf{CC3:} $z=z'$ and $k = k'$.
\end{itemize}
Consider the  weight of each vertex $v_{k,z,n,j}$ is defined as $w(v)=T_lC_kD_n\alpha f^2_k+  \frac{p_n\mathtt D_n}{R^n_{k,z}}+T_lC_kD_j\alpha f^2_k+  \frac{p_j\mathtt D_j}{R^j_{k,z}}$.  Thus, the vertex' weight becomes small when the data transmission time from the represented IoT devices is small as well as the local computation time at the represented F-AP is small. This yields to a smart
scheduling of IoT devices representing the corresponding vertex with a smaller weight, which in turn minimizes
the energy consumption. Therefore, any minimal independent set in NOMA-coordinated graph represents a set of NOMA clusters that satisfies the following criterion: 1) each IoT device in the set is scheduled to only one F-AP and one RRB, 2) each RRB identified by the vertices
in a minimal independent set represents a NOMA cluster of two IoT devices, and 3) the total number of scheduled IoT devices at each F-AP is not larger than $U$.

The following theorem characterizes the solution of allocating IoT devices to the RRBs across all F-APs  such that the total energy consumption is minimized.

\textit{\textbf{Theorem 1:}}
\textit{
	The IoT device coordinated scheduling problem $\mathcal P_{9}$ is equivalent to MWIS problem over the NOMA-coordinated graph, wherein the weight of a vertex $v_{k,z,n,j}$ is given by
	\begin{align} \label{eq39}
	w(v) = \alpha f^2_kT_lC_k\left(D_n+D_j\right)+ \frac{p_n\mathtt D_n}{R^n_{k,z}}+ \frac{p_j\mathtt D_j}{R^j_{k,z}}.
	\end{align}
	The set of scheduled IoT devices to the $z$-th	RRB in the $k$-th F-AP is obtained by combining the vertices of the MWIS $\mathbf{I}$ in the NOMA-coordinated  graph.}

\begin{proof}
 This theorem can be proved by demonstrating the
following facts. The first fact establishes the equivalency between $\mathcal P_{9}$ and MWIS problems. Specifically, using $\mathcal{G}_\text{NOMA}$, $\mathcal P_{9}$ is similar to
MWIS problems. In  MWIS problems, two vertices must be nonadjacent in the graph, and similarly, in problem $\mathcal P_{9}$, two NOMA clusters cannot be allocated with the same RRB or contain at least one IoT device. Afterward, the weight of each vertex is set to be the minimum energy consumption contribution of the corresponding NOMA cluster to the network. Therefore, the MWIS is a
feasible solution with the minimum energy consumption, i.e., the MWIS is the feasible solution to
$\mathcal P_{9}$.  To finalize the proof, we now prove that the weight of the MWIS is the objective function in $\mathcal P_{9}$ to be minimized. Let $\mathbf{I}=\{v_1,v_2,\ \cdots, \, v_{|\mathbf{I}|}\}$, $v\in \mathcal{G}_\text{NOMA}$. Let a vertex $v \in \mathcal{V}$ is associated with 2-IoT devices NOMA cluster $(n,j)$. The weight of the MWIS over all the vertices that are representing the corresponding NOMA clusters over all RRBs/F-APs can be written as 
\begin{align} 
\begin {split}
w(\mathbf{I})&= \sum\limits _{v\in \mathbf{I}}w(v)\\&= \sum\limits _{\mathbf{k}\in \mathcal{K}} \sum\limits _{z\in \mathcal{Z}}\left(\alpha f^2_kT_lC_k\left(D_n+D_j\right)+ \frac{p_n\mathtt D_n}{R^n_{k,z}}+ \frac{p_j\mathtt D_j}{R^j_{k,z}}\right).
\end{split}
\end{align}

Therefore, the problem of minimizing the energy consumption $\mathcal P_{9}$ is equivalent to the MWIS problem among the minimal sets in the NOMA coordinated graph.
\end{proof}

\textit{2) \textbf{Heuristic Solution:}} MWIS problems are NP-hard problems, where the required complexity of solving these problems optimally requires an exhaustive search of  $|\mathcal{V}|^2.2^{|\mathcal{V}|}$ complexity where $\mathcal{V}$ is the set of vertices of graph $\mathcal{G}_\text{NOMA}$.  However, MWIS problems can be heuristically solved with a reduced complexity of  $b^{|\mathcal{V}|}$ where $b$ is the complexity constant \cite{25, 26}. Thus, the MWIS problem can be solved effectively using a low-complexity  heuristic solution.

Let $w(v)$ be the raw weight of vertex $v$ in the NOMA coordinated graph as expressed in \eref{eq39}. The modified weight $\tilde{w}(v)$ of
vertex $v$ can be defined as
\begin{equation} \label{eq41}
\tilde{w}(v)= w(v)\sum_{v^\prime\in\mathcal{V}_v}w({v^\prime}),
\end{equation}
where $\mathcal{V}_v$ is the set of vertices not connected to vertex $v$ by transmission conflict edges. The appropriate design of the weights shows that $\tilde{w}_v$ reflects the contribution of the vertex to the network as it has a small raw weight and non-adjacent to a large number of vertices induced by users with small raw weight. The two-phase scenario
of IoT device coordinated scheduling and F-AP's computation frequency allocation is presented in Algorithm \ref{alg3}.

\begin{algorithm}[t!]
	
	\textbf{Data:} $\mathcal{N}, \mathcal{K}, \mathcal{Z}$, $p_\text{max}$, $f_k^\text{min}$, $f_k^\text{max}$, $T_q$, $C_kD_k$, $d_k$, and $G^n_{k,z}$,  $(n,k,z)\in\mathcal{N}\times\mathcal{K}\times\mathcal{Z}$.\;
	\textbf{Phase I: IoT device coordinated scheduling}\;
	\begin{itemize}
		\item Initialize $\mathcal G_\text{NOMA}=\emptyset$, $\mathbf I=\emptyset$.\;
		\item Construct $\mathcal G_\text{NOMA}$ using \sref{MT-B}.\;
		\item For each $v\in \mathcal G_\text{NOMA}$, calculate $w(v)$ and $\tilde{w}(v)$ using \eref{eq39}, \eref{eq41}, respectively\;
		\item Solve the MWIS problem in $\mathcal{G}_\text{NOMA}$ to find $\mathbf{I}$ as follows:\;
		
		\While{$\mathcal{G} \neq\ \emptyset$}{
		$v^\ast = \min_{v\in\mathcal{G}_\text{NOMA}} \{{w(v)}\}$\;
			Set $\mathbf{I}$ = $\mathbf{I}\ \cup v^{\ast}$ and set $\mathcal{G}_\text{NOMA}=\mathcal{G}_\text{NOMA}(v^{\ast })$\;  Continue only with vertices not linked to $v^{\ast}$ in $\mathcal{G}_\text{NOMA}$\;}
	
		\item Output: The MWIS $\mathbf I$.\;
		\item For the resulting $\mathbf I$, solve the power allocation problem $\mathcal P_{10}$.
		\item Continue iterating between finding $\mathbf I$ and solving $\mathcal P_{10}$ until convergence. 
	\end{itemize}
	\textbf{Phase II: F-APs' computation frequency allocation}\;
	\begin{itemize}
	\item \textbf{Solve $\mathcal P_{8}$ for the resulting $\mathbf I$ and power allocation}.\;
	\For{$v=\{v_1,v_2,\ \cdots, |\mathbf I|\}$}{
	 	Calculate $\hat f_k=\frac{T_lC_k B_k}{T_q-\max_{n\in \mathcal N_k}\left(\frac{\mathtt D_n}{R_{k,z}^{*n}}\right)-\frac{d_k}{R_{fh}}}$, $\forall (k,z,n)$ is associated with $v$\;
       Calculate $f_k$ according to \textit{Lemma 3}

   }
	 \item  Execute phases 1 and II until convergence or a maximum number of iteration is reached.
	  \item Obtain $\mathbf I$ and $\mathbf f_K$.
	\end{itemize}
	\caption{Coordinated Scheduling Algorithm} \label{alg3}
\end{algorithm}

\subsection{Complexity Analysis}
The computational complexity of Algorithm \ref{alg3} is dominated by the required complexity of generating feasible NOMA clusters and connecting  the generated vertices. To generate all the vertices using the low-complexity graph method, a total of $\mathcal{O}\left(KZ\binom{|\mathcal N_k|}{2}\right)$ computational complexity is required. On the other hand, the required complexity of connecting the generating vertices, i.e., the required complexity of finding neighborhood  of the generated vertices is $\mathcal{O}\left(\left(KZ\binom{|\mathcal N_k|}{2}\right)^2\right)$. Therefore, the overall computational complexity of Algorithm \ref{alg3} is obtained as $\mathcal{O}\left(NKZ+\left(KZ\binom{|\mathcal N_k|}{2}\right)^2\right)\approx \mathcal{O}\left(\left(KZ\binom{|\mathcal N_k|}{2}\right)^2\right)$.\ignore{ Accordingly, Algorithm \ref{alg2} requires polynomial computational complexity.}

\section{Numerical Results}\label{NR}
\ignore{This section presents selected simulation results that compare the energy consumption and FL times performances of our proposed two
schemes with baseline algorithms.}
\subsection{Simulation Setting and Schemes Under Consideration}
In our simulations, we consider a hexagonal cell of radius $1500$ m where F-APs and CS have fixed locations and IoT devices are distributed randomly within the cell. The CS is
located at the cell center. The channel model for IoT device-F-AP transmissions follows the standard path-loss model, which consists of three components: 1) path-loss of $128.1+37.6\log_{10}(\text{dis.[km]})$; 2) log-normal shadowing with $4$ dB standard deviation; and 3) Rayleigh channel fading with zero-mean and unit variance. The noise power, F-AP's power, and maximum' IoT device power are assumed to be $-174$ dBm/Hz and $q_k=p_\text{max}=3$ W, respectively \cite{FCR11}. The weighting factor $V$ is set to $0.3$. The total number of global and local FL iterations are calculated as $T_g=\frac{2\beta^2}{(2\vartheta-\beta \eta)\vartheta \eta}\ln(1/\epsilon_g)$, $T_l=\frac{2}{(2-\delta \beta)\delta \vartheta}\ln(1/\epsilon_l)$, respectively, with $\beta=4, \eta=1/3, \delta=1/4, \vartheta=2, \epsilon_g=\epsilon_l=10^{-3}$ \cite{7a}. The FL time threshold $T_q$ is $1$ second \cite{7a}. The bandwidth of each RRB is $20$ MHz. Unless otherwise stated, we set the numbers of F-APs and RRBs to $9$ and $4$, respectively. The fronthaul capacity $R_{fh}$ is set to 150 Mbit/s. For each IoT device and each F-AP, the number of data samples $D_n$ is randomly chosen from $800$ to $1000$. Other parameters are summarized in Table \ref{table_1}. To assess the performance of our proposed scenarios, we simulate various scenarios with different number of IoT devices $N$, data size $\mathtt D_n$, number of RRBs $Z$, number of data samples $D_n$, computation frequency allocation $f_n$, $f_k$, and parameter data size. For the sake of comparison, our
proposed schemes are compared with the following baseline schemes.
\begin{itemize}
	\item \textbf{Power-only:} This scheme minimizes the energy consumption by optimizing the power level of IoT devices
	and fixing the computation frequency allocation to its maximum value. 
	\item \textbf{Computation frequency-only:} This scheme, denoted by \textit{CPU-only}, minimizes the energy consumption by optimizing the computation frequency allocation
	and fixing the power level to its maximum value. 
	\item \textbf{Fixed}: This scheme employs random IoT device scheduling and fixes both 	the computation frequency allocation and transmission power to their maximum values. 
\end{itemize}

\begin{table}[t!]
	\renewcommand{\arraystretch}{0.9}
	\caption{Simulation Parameters}
	\label{table_1}
	\centering
	\begin{tabular}{p{5.4cm}| p{4.3cm}}
		\hline
		\hline
		
		\textbf{Parameter} & \textbf{Value}\\
		\hline 
		Circle radius of F-AP’s service area $\mathtt R$ & $500$ m\\
		\hline
		learning local  parameter size, $d_n$, $d_k$ & $[5-10]$ Kbit \cite{7a}\\
		\hline
		IoT device data size, $\mathtt D_n$ & $[0.5-1]$ Mbit\\
		\hline
		IoT device processing density, $C_n$ & $[600-800]$ \cite{7a} \\
		\hline
		F-AP processing density, $C_k$ & $[1000-1500]$ \\
		\hline
		IoT device computation frequency, $f_{n}$ & $[0.0003-1]$ G cycles/s \cite{7a}\\
		\hline
		F-AP computation frequency, $f_{k}$ & $[0.0005-5]$ G cycles/s \\
		\hline
		CPU architecture based parameter, $\alpha$ & $10^{-28}$ \cite{14}\\
		\hline
		
	\end{tabular}
\end{table}

\vspace*{-0.2cm}
\subsection{Simulation Results and Discussions}
We adopt two performance metrics as follows: (i) the
\textit{energy consumption}  that represents the
objective in $\mathcal P_1$ for the IoT device local learning scenario and $\mathcal P_2$ for
the F-AP local learning scenario, and (ii) the \textit{FL time} as expressed in \eref{FL_1}.

\begin{figure}[t!]
	\centering
	\begin{minipage}{0.494\textwidth}
		\centering
		\includegraphics[width=0.75\textwidth]{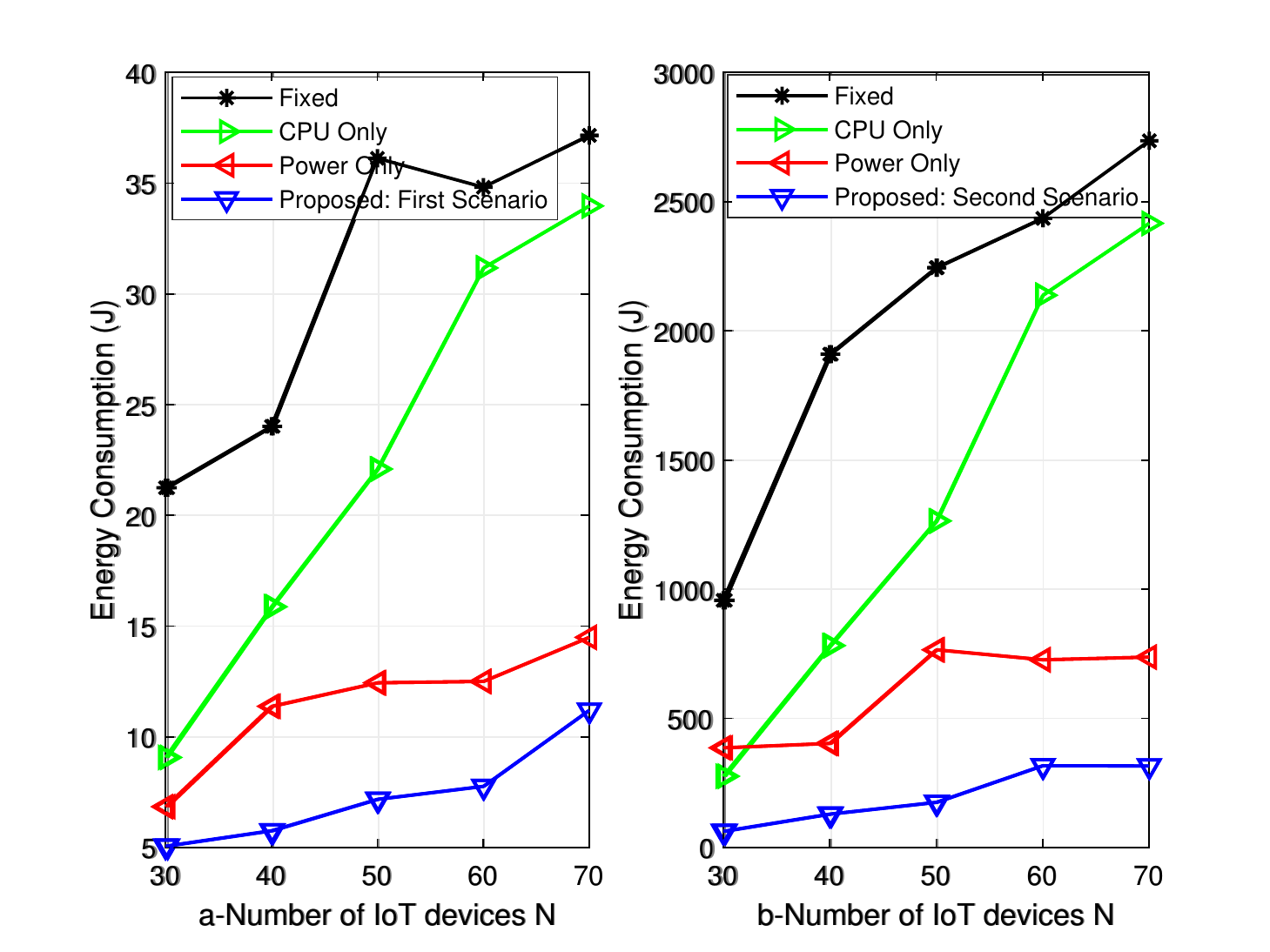} 
		\caption{Energy consumption vs. number of IoT devices $N$ for $K = 9$ and $Z=4$.}
		\label{fig3}
	\end{minipage}\hfill
	\begin{minipage}{0.494\textwidth}
		\centering
		\includegraphics[width=0.75\textwidth]{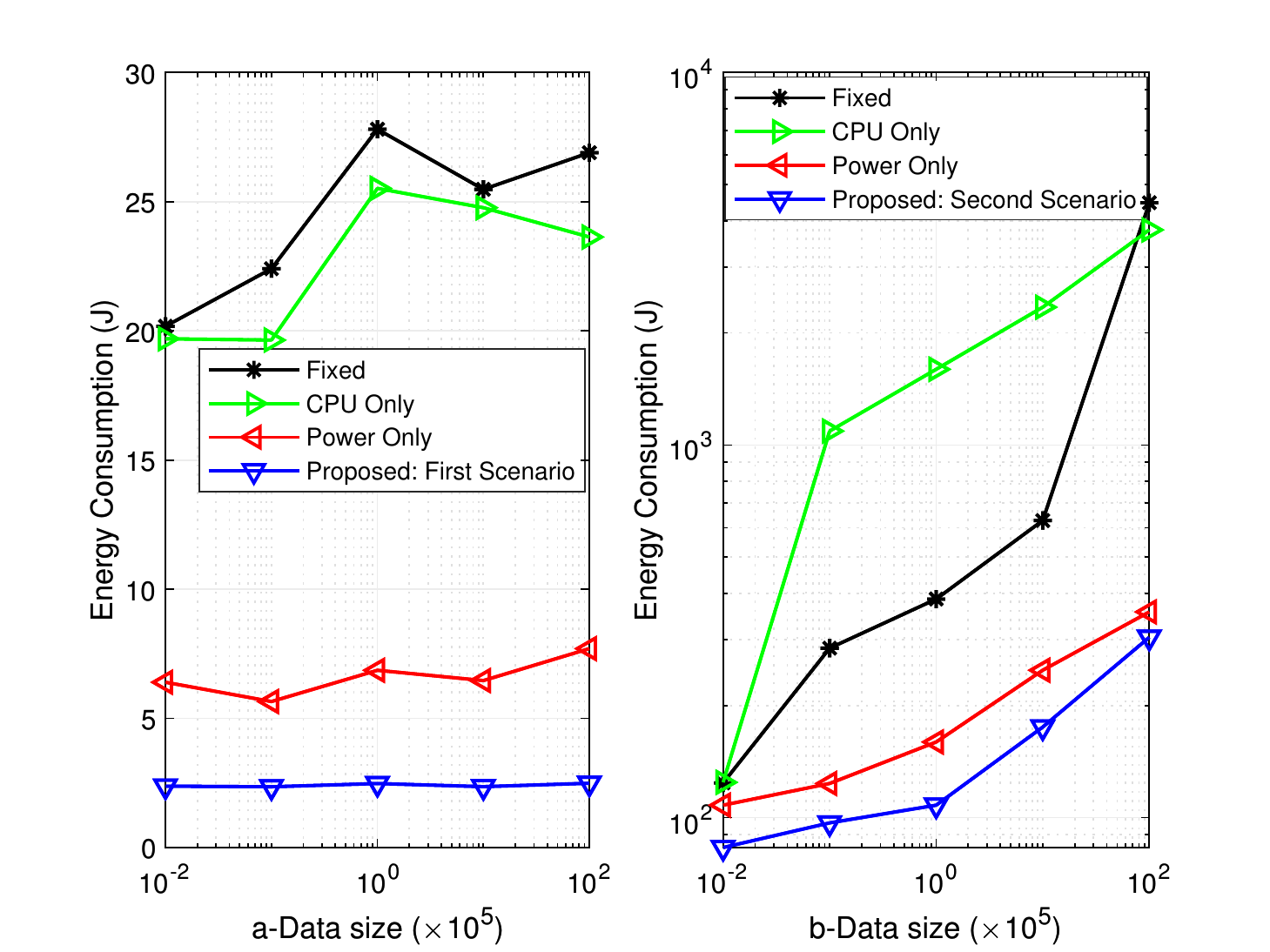} 
		\caption{Energy consumption vs. data size for $N=50$, $K = 9$ and $Z=4$.}
		\label{fig4}
	\end{minipage}\hfill
\end{figure}

\textit{1) Consumption energy performance:}
In Figs. \ref{fig3}-a and \ref{fig3}-b, we plot the energy consumption versus the number of IoT devices for the first and second scenarios, respectively. Our proposed schemes have 
the following two attributes. First, they judiciously schedule IoT devices
to F-APs/RRBs, adapt the transmission rate of each IoT device, and optimize the transmission power of each IoT device. Second, our proposed schemes efficiently optimize the computation frequency allocation of IoT devices and F-APs. Leveraging these two attributes, our proposed schemes significantly reduce the energy consumption compared to the benchmark schemes, as depicted from both  Figs. \ref{fig3}-a and \ref{fig3}-b.
 In particular, the \textit{power-only} scheme selects the maximum computation frequency allocation for each IoT device and each F-AP. Consequently, the \textit{power-only} scheme results in  higher energy consumption for local learning, and it increases the energy consumption of the system in both scenarios. 
 The \textit{CPU-only} scheme ignores the power optimization that leads to more interference among the IoT devices and increased offloading transmission time. As a result, the \textit{CPU-only}  scheme leads to a high energy consumption.  Finally, the \textit{fixed} scheme has the most energy consumption
 because it chooses the maximum CPU frequency and
 transmission power.  Accordingly, from an energy consumption perspective, it is inefficient to offload data to F-APs while ignoring the power allocation and employing random IoT device scheduling to F-APs/RRBs. 


In Figs. \ref{fig4}-a and \ref{fig4}-b, we plot the energy consumption versus the
data size $\mathtt D_n$ for the first and second scenarios, respectively. When the data size is small (around $1$ Kbit), both proposed schemes work superior in terms of minimizing the energy consumption. When the data size is nearly $10$ Mbit, the energy consumption performance of our proposed first scheme does not change much and has a performance of $3$ J. This is becasue the IoT devices perform local learning on the data and offload the local learning parameters only. However, when the data size changes from $1$ Kbit to $10$ Mbit, the energy consumption performance of the proposed second scheme changes form $1$ J to around $303$ J. Therefore, our  proposed second scheme  consumes more energy when the data size increases.  Accordingly, from an energy consumption perspective, learning at the IoT devices as in the first proposed scenario is more efficient, especially for  large numbers of IoT devices and large data sizes.

\begin{figure}[t!]
	\centering
	\begin{minipage}{0.494\textwidth}
		\centering
		\includegraphics[width=0.75\textwidth]{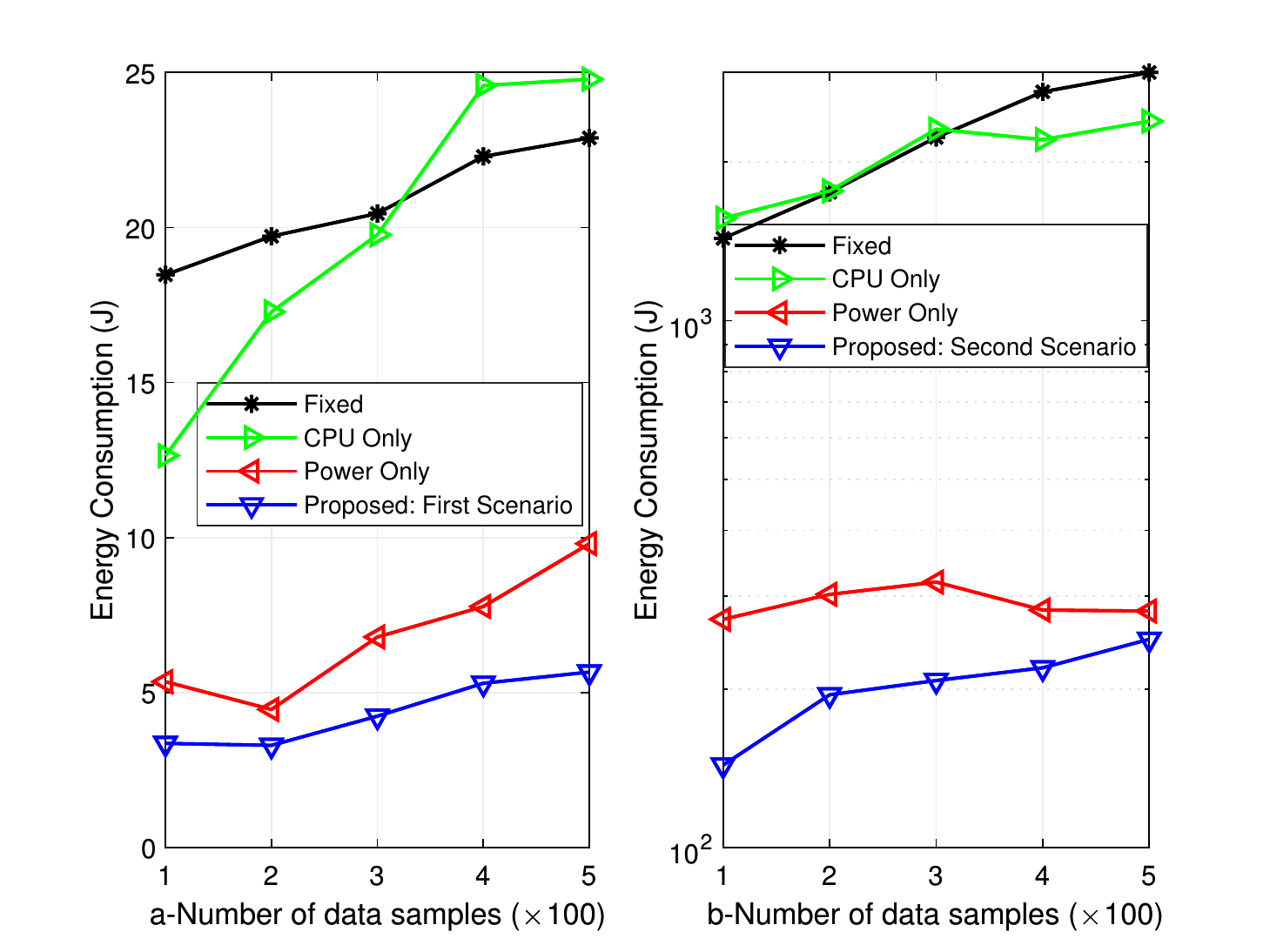} 
		\caption{Energy consumption vs. number of data samples for $N=50$, $K = 9$ and $Z=4$.}
		\label{fig5}
	\end{minipage}\hfill
	\begin{minipage}{0.494\textwidth}
		\centering
		\includegraphics[width=0.75\textwidth]{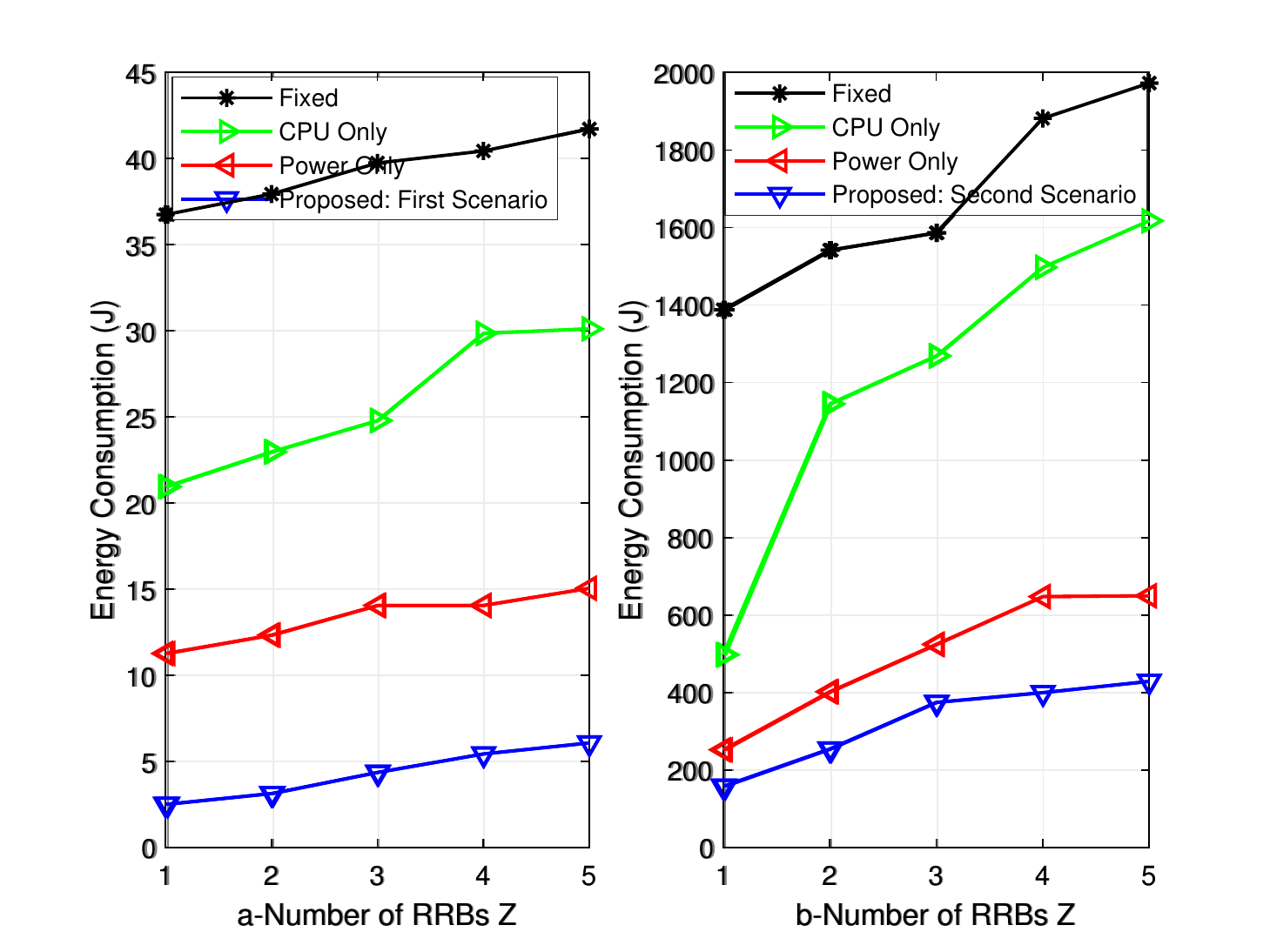} 
		\caption{Energy consumption vs. number of RRBs $Z$ for $N=50$ and $K = 9$.}
		\label{fig6}
	\end{minipage}\hfill
\end{figure}

In Figs. \ref{fig5}-a and \ref{fig5}-b, we show the energy consumption
versus the number of data samples $D_n$ for the first and second scenarios, respectively. The number of data samples affects the CPU-related energy consumption. Similar to our discussions for Fig. \ref{fig3}, the \textit{CPU-only} and \textit{fixed} schemes severely degrades the energy consumption performance. Specifically,
the energy consumption of  the \textit{fixed} scheme is increased with the number of data samples. However,  the energy consumption  of the \textit{power-only} and our proposed schemes do not significantly
change, e.g., see Fig. \ref{fig5}-(a). Since the energy consumption in the second scenario is dominated by data offloading to F-APs, both \textit{CPU-only} and \textit{fixed} schemes consume high energy as can be seen from  Fig. \ref{fig5}-(b). Using the optimized resource allocations, our proposed schemes incur the least energy consumption for both small and large data samples.

In Figs. \ref{fig6}-a and \ref{fig6}-b, we plot the energy consumption
versus the number of RRBs $Z$ for the first and second scenarios, respectively. As can be seen, the consumed energy of all schemes
are increased with the increase in the number of RRBs. This is due to the fact that as the number of RRBs
increases, more IoT devices are scheduled, which in turn increases the energy consumption. Specifically, when $Z=1$, the maximum number of accommodated IoT devices by the F-APs is $2ZK=2\times 1\times 9=18$, thus the consumed energy of all schemes is low. As the number of RRBs is increased,
the energy consumption of all the schemes is increased. This can be explained by
the fact that when the number of RRBs goes beyond $2$, no more IoT devices can be accommodated. Thus, the consumed energy of all schemes do not change much. For a fair comparison, we consider that all the schemes serve the same set of IoT devices in  the available RRBs of a given F-AP.  The proposed schemes, however, benefit from optimizing the transmit power and computation frequency allocation. Essentially, the proposed schemes achieve reduced energy consumption  compared to the benchmark schemes.

\begin{figure}[t!]
	\centering
	\begin{minipage}{0.494\textwidth}
		\centering
		\includegraphics[width=0.75\textwidth]{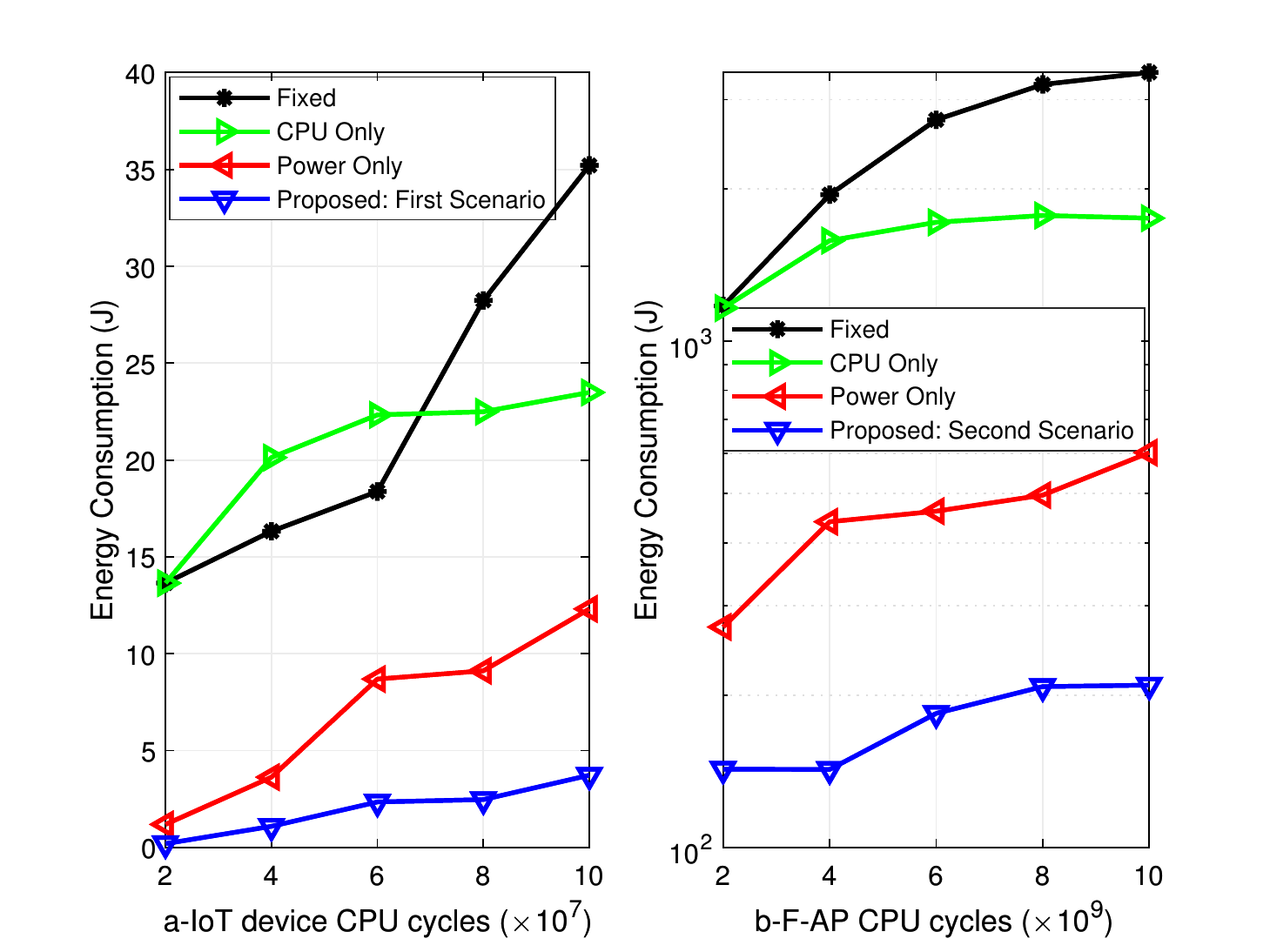} 
		\caption{Energy consumption vs. computation frequency for $N=50$, $K = 9$ and $Z=4$.}
		\label{fig7}
	\end{minipage}\hfill
	\begin{minipage}{0.494\textwidth}
		\centering
		\includegraphics[width=0.75\textwidth]{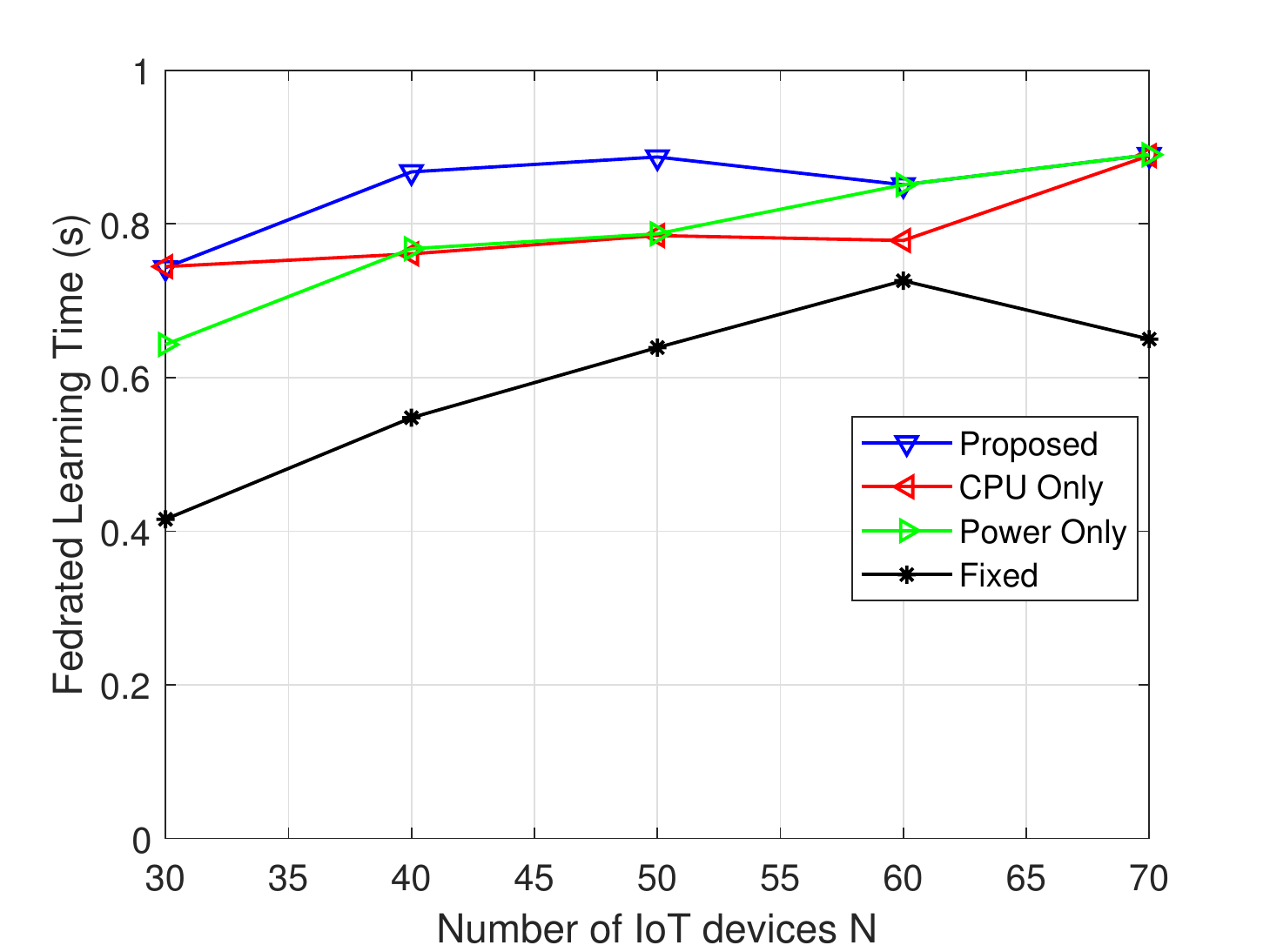} 
		\caption{Federated learning time vs. number of IoT devices $N$ for $K = 9$ and $Z=4$.}
		\label{fig8}
	\end{minipage}\hfill
\end{figure}

In Figs. \ref{fig7}-a and \ref{fig7}-a , we show the energy consumption
versus the number of CPU cycles for the first and second scenarios, respectively. The number of CPU cycles ranges from $2\times10^7$ to $10\times10^7$ in Fig. \ref{fig7}-(a) and from $2\times10^9$ to $10\times10^9$ in Fig. \ref{fig7}-(b). The number of CPU cycles determines the energy consumption.
Hence, the energy consumption of both \textit{fixed} and \textit{power-only} schemes, that fix the computation frequency allocation
at the highest value, is considerably   increased with the increasing number of computation cycles. On the other hand, the energy consumption of both
\textit{CPU-only} and proposed schemes with adjustable CPU frequencies do not change much as shown in Figs. \ref{fig7}-a and \ref{fig7}-b. As expected, using both transmit power and computation frequency allocation, our proposed schemes incur the least energy consumption for both small and large numbers of CPU cycles.

\begin{figure}[t!]
	\centering
	\begin{minipage}{0.494\textwidth}
		\centering
		\includegraphics[width=0.75\textwidth]{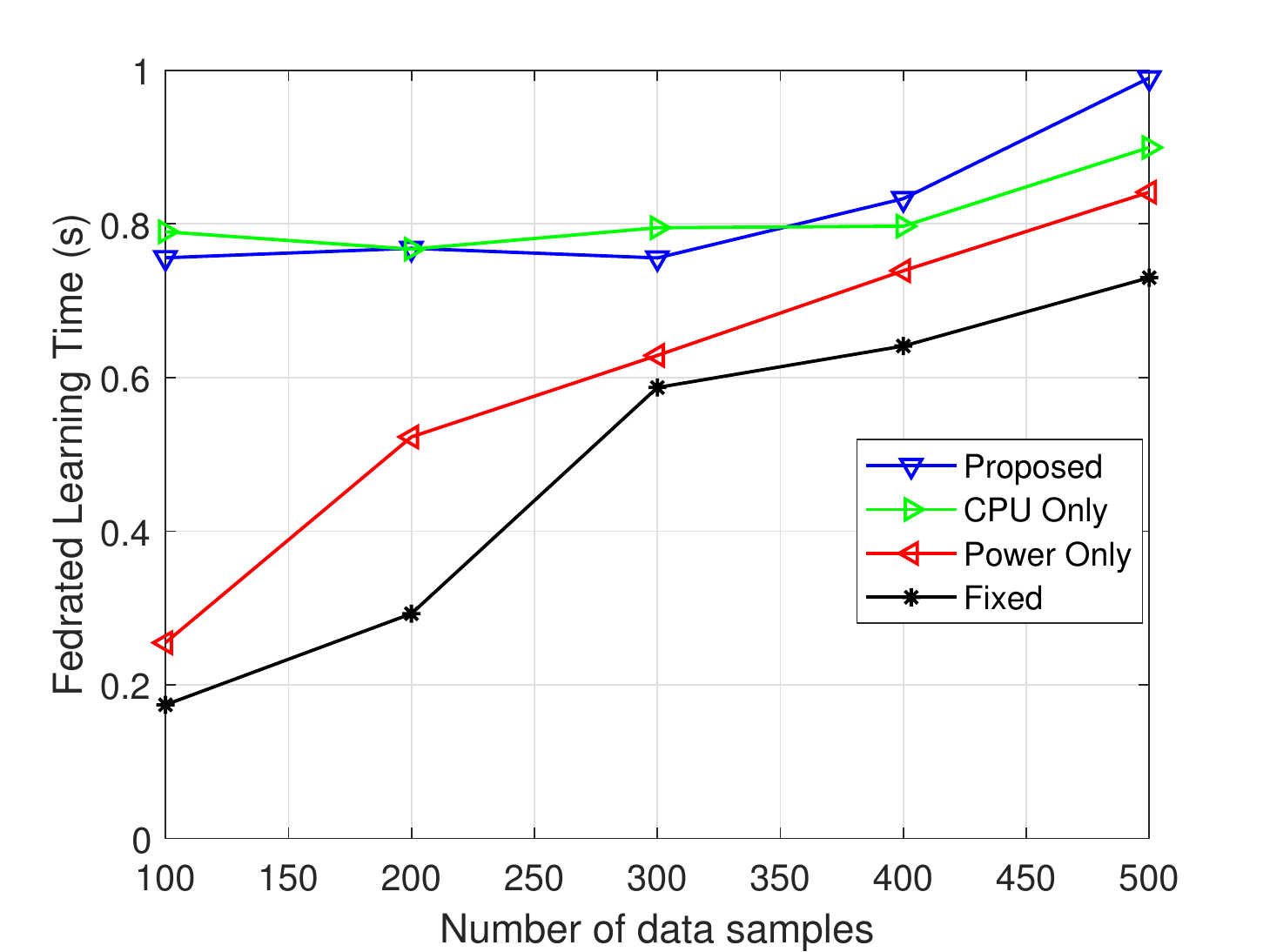} 
		\caption{Federated learning time vs. number of data samples $D_n$ for $K = 9$ and $Z=4$.}
		\label{fig9}
	\end{minipage}\hfill
	\begin{minipage}{0.494\textwidth}
		\centering
		\includegraphics[width=0.75\textwidth]{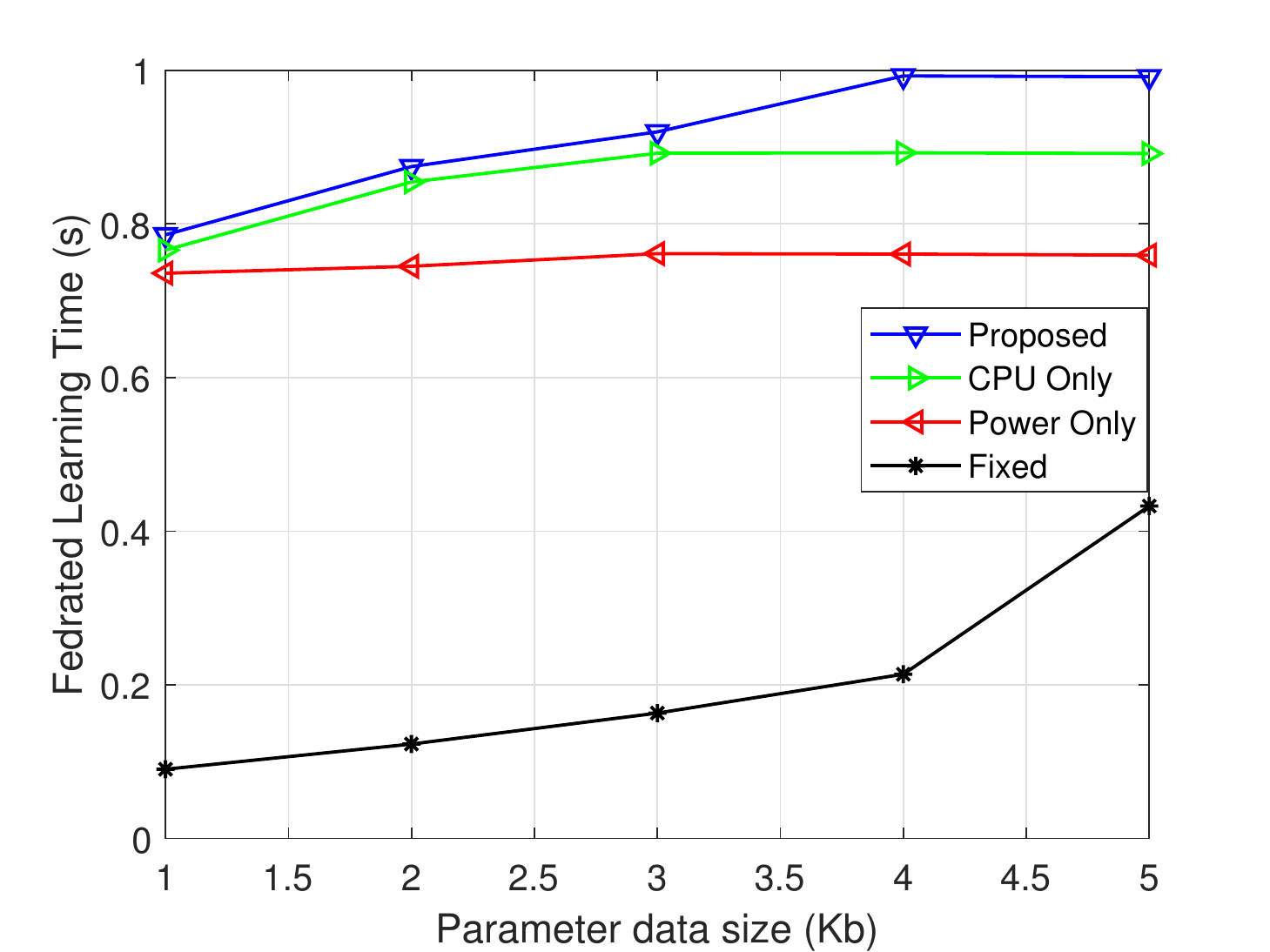} 
		\caption{Federated learning time vs. parameter data size $d_n$ for $K = 9$ and $Z=4$.}
		\label{fig10}
	\end{minipage}\hfill
\end{figure}

\ignore{
\begin{figure}[t!]
	\centering
	\begin{subfigure}[t]{0.28\textwidth}
		\centerline{\includegraphics[width=1.11\linewidth]{fig1FL.eps}}
		\label{MCT}
	\end{subfigure}%
	~
	\begin{subfigure}[t]{0.28\textwidth}
		\centerline{\includegraphics[width=1.13\linewidth]{fig2FL.eps}}
		\label{NCT}
	\end{subfigure}
	~
	\begin{subfigure}[t]{0.28\textwidth}
		\centerline{\includegraphics[width=1.13\linewidth]{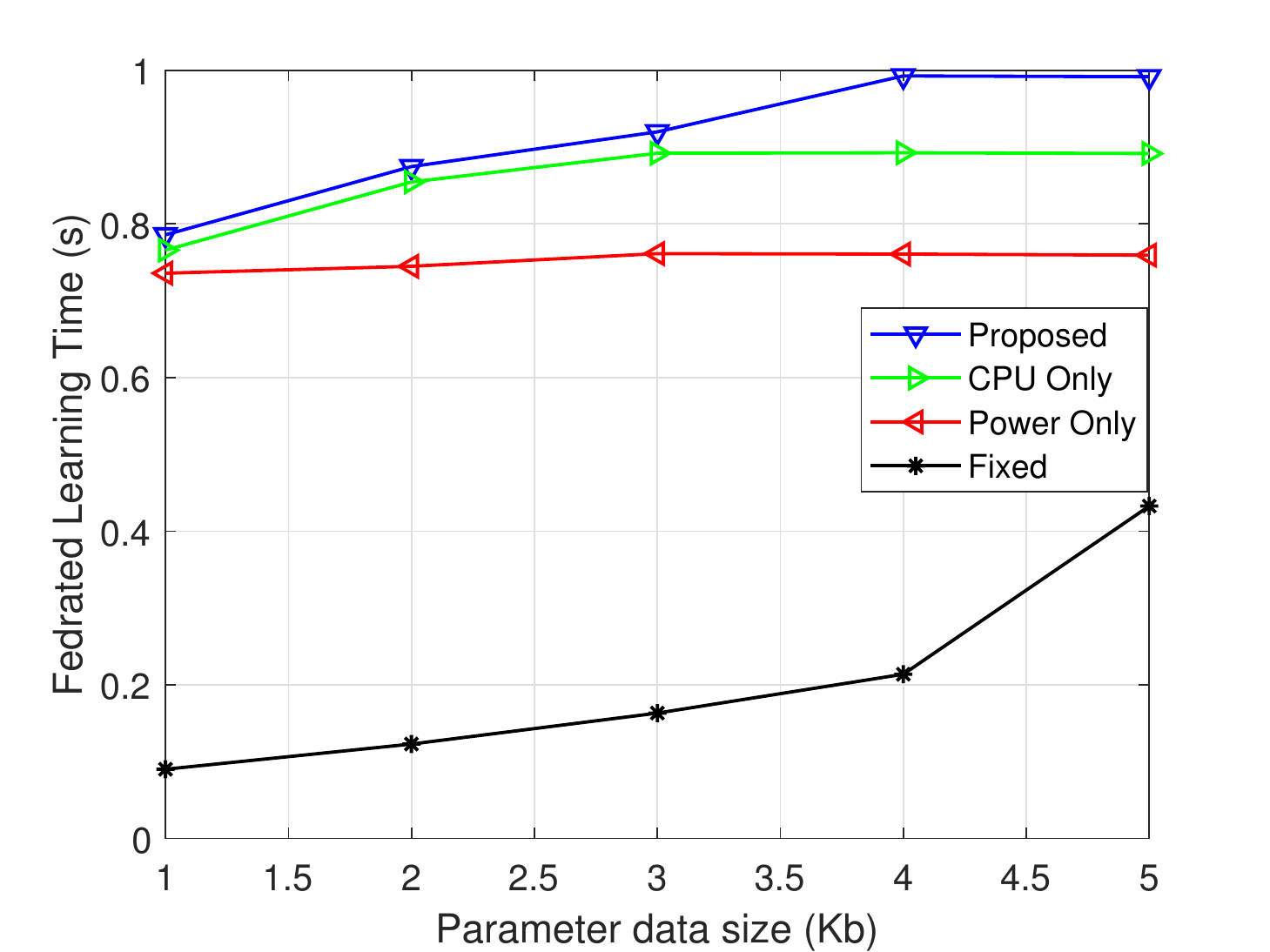}}
		\label{MCT}
	\end{subfigure}%
	
	\caption{FL time versus: (a) number of IoT devices $N$, (b) number of data samples $D_n$, and (c) parameter data size $d_n$.}
	\label{figFL}
\end{figure}}

\textit{2) FL time performance:} In Fig. \ref{fig8}, Fig. \ref{fig9}, and Fig. \ref{fig10}, we plot the federated learning time versus: (a) number of IoT devices $N$, (b) number of data samples $D_n$, and (c) parameter data size $d_n$, respectively. First, it is clear that the FL time depends on the transmission time and the computation learning time of IoT devices. Since the local learning parameters have small size, the transmission
time for offloading such parameters to F-APs/CS requires smaller portion of the overall  FL time compared with the computation training time. Consequently, the FL time is dominated by the computation training time. As can be seen from Figs. \ref{fig8}, \ref{fig9}, \ref{fig10}, \textit{fixed} scheme, that chooses the maximum CPU frequency, effectively minimizes the FL time  at the cost of consuming the most energy as shown in Fig. \ref{fig3} to Fig. \ref{fig7}. Our proposed first scheme that considers IoT device local learning, denoted by \textit{proposed}, adjusts the CPU frequency and power transmissions so that it effectively minimizes the consumed energy within the FL time of $1$ second. In Figs. \ref{fig8}, \ref{fig10}, the FL time of all algorithms
does not change much with the number
of IoT devices and local parameter size. This is because the FL time
is mainly controlled by the longest local training time of one IoT device, which does not significantly change  when  the number of IoT device and the local parameter size are increased.

Finally, we provide some observations from our presented simulation results as follows. First, although the \textit{fixed} scheme performs fairly well in terms of reducing the FL time, it exhibits a poor energy consumption performance, which is impractical. Thus, it
only serves as a benchmark scheme in this work.  Second, it is advantageous to optimize the computation frequency allocation of the IoT devices and the F-APs as in the \textit{CPU-only} scheme. However, it is inefficient to ignore the power optimization that significantly impacts the energy consumption of the system. Third, the \textit{power-only} scheme works well in terms if reducing the energy consumption; however, its performance is degraded since it uses the maximum CPU of each IoT and each F-AP. Fourth, our proposed schemes strike a balance
between the aforementioned aspects by judiciously scheduling IoT devices
to F-APs/RRBs, adapting the transmission rate of each IoT device, and optimizing the transmission power of each IoT device. Furthermore, our proposed schemes efficiently optimize the computation frequency allocation of IoT devices and F-APs. Finally,  as the data size increases,
the energy consumption performance of the second proposed scheme degrades. This is because as the data size increases, the transmission time for offloading IoT devices' data to F-APs is significantly increased. Thus, the energy efficiency of our proposed first scheme becomes more pronounced compared to our second proposed scheme.  

\section{Conclusion} \label{C}
In this paper, we  investigated the resource allocation strategy to minimize the energy consumption  for performing FL in an integrated FCC-enabled IoT network subject to FL time constraint. Specifically, we considered two scenarios for training the local models, and for both scenarios, we proposed joint optimization
of computation frequency allocation, IoT device scheduling, and  transmission power control of  network edge devices. Leveraging  graph theory, we proposed efficient iterative schemes. The presented numerical results revealed that the proposed schemes substantially reduce the energy consumption  compared to the baseline solutions, at the cost of small increase of FL learning time.  The presented simulation results interestingly revealed that for a large number of IoT devices and large data sizes, it is more energy efficient to train the local models at the IoT devices instead of the F-APs.



\end{document}